\documentclass[preprint]{aastex61}
\usepackage{latexsym,bm}
\usepackage{subfigure}
\usepackage{graphicx}
\usepackage{color}
\usepackage{natbib}

\shorttitle{Asteroseismology of VX\ Hya}
\shortauthors{Xue et al.}

\begin{document}

\title{Asteroseismology of the Double-Mode High-Amplitude $\delta$ Scuti Star VX\ Hydrae}

\correspondingauthor{Jian-Ning Fu}
\email{jnfu@bnu.edu.cn}

\author{Hui-Fang Xue}
\affil{Department of Astronomy, Beijing Normal University, Beijing 100875, People's Republic of China}

\author{Jian-Ning Fu}
\affil{Department of Astronomy, Beijing Normal University, Beijing 100875, People's Republic of China}

\author{L. Fox-Machado}
\affil{Observatorio Astron\'omico Nacional, Instituto de Astronom\'{\i}a, Universidad Nacional Aut\'onoma de M\'exico, Ap. P. 877, Ensenada, BC22860, M\'exico}

\author{Jian-Rong Shi}
\affil{Key Laboratory of Optical Astronomy, National Astronomical Observatories, Chinese Academy of Sciences, Beijing 100012, People's Republic of China}
\affil{University of Chinese Academy of Sciences, Beijing 100049, People's Republic of China}

\author{Yu-Tao Zhou}
\affil{Key Laboratory of Optical Astronomy, National Astronomical Observatories, Chinese Academy of Sciences, Beijing 100012, People's Republic of China}
\affil{University of Chinese Academy of Sciences, Beijing 100049, People's Republic of China}

\author{Jun-Bo Zhang}
\affil{Key Laboratory of Optical Astronomy, National Astronomical Observatories, Chinese Academy of Sciences, Beijing 100012, People's Republic of China}

\author{R. Michel}
\affil{Observatorio Astron\'omico Nacional, Instituto de Astronom\'{\i}a, Universidad Nacional Aut\'onoma de M\'exico, Ap. P. 877, Ensenada, BC22860, M\'exico}

\author{Hong-Liang Yan}
\affil{Key Laboratory of Optical Astronomy, National Astronomical Observatories, Chinese Academy of Sciences, Beijing 100012, People's Republic of China}

\author{Jia-Shu Niu}
\affil{Institute of Theoretical Physics, Shanxi University, Taiyuan, 030006, People's Republic of China}
\affil{Key Laboratory of Theoretical Physics, Institute of Theoretical Physics, Chinese Academy of Sciences, Beijing, 100190, People's Republic of China}
\affil{School of Physical Sciences, University of Chinese Academy of Sciences, No.19A Yuquan Road, Beijing 100049, People's Republic of China}

\author{Wei-Kai Zong}
\affil{Department of Astronomy, Beijing Normal University, Beijing 100875, People's Republic of China}

\author{Jie Su}
\affiliation{Yunnan Observatories, Chinese Academy of Sciences, Kunming 650216, China}

\author{A. Castro}
\affil{Observatorio Astron\'omico Nacional, Instituto de Astronom\'{\i}a, Universidad Nacional Aut\'onoma de M\'exico, Ap. P. 877, Ensenada, BC22860, M\'exico}
\affil{Physics and Astronomy, University of Southampton, Southampton S017 1BJ, UK}

\author{C. Ayala-Loera}
\affil{Observatorio Astron\'omico Nacional, Instituto de Astronom\'{\i}a, Universidad Nacional Aut\'onoma de M\'exico, Ap. P. 877, Ensenada, BC22860, M\'exico}

\author{Altamirano-D\'evora L.}
\affil{Observatorio Astron\'omico Nacional, Instituto de Astronom\'{\i}a, Universidad Nacional Aut\'onoma de M\'exico, Ap. P. 877, Ensenada, BC22860, M\'exico}

\begin{abstract}
  Bi-site time-series photometric and high-resolution spectroscopic observations were made for the double-mode high-amplitude $\delta$ Scuti star VX\ Hya. The fundamental frequency $f_{0}=4.4763\ \rm{c\ days^{-1}}$, the first overtone $f_{1}=5.7897\ \rm{c\ days^{-1}}$ and 23 harmonics and linear combinations of $f_{0}$ and $f_{1}$ are detected by pulsation analysis. From the spectroscopic data, we get $\rm{[Fe/H] = -0.2\pm0.1\ dex}$. The period change rate of the fundamental mode is obtained by using the Fourier-phase diagram method, which gives the value of $(1/P_{0})(dP_{0}/dt)=(1.81\pm0.09) \times 10^{-7}\ \rm{yr}^{-1}$. With these results from the observations, we perform theoretical explorations with the stellar evolution code MESA, and constrain the models by fitting $f_{0}$, $f_{1}$, and $(1/P_{0})(dP_{0}/dt)$ within $3\sigma$ deviations. The results show that the period change of VX\ Hya could be ascribed to the evolutionary effect. The stellar parameters of VX\ Hya could be derived as: the mass $2.385\pm0.025\ M_{\odot}$, the luminosity $\log(L/L_{\odot})=1.93\pm0.02$ and the age $(4.43\pm0.13)\times 10^8$ years. VX\ Hya is found to locate at the post-main-sequence stage with a helium core and a hydrogen-burning shell on the H${-}$R diagram.
\end{abstract}


\section{introduction}

$\delta$ Scuti stars are a class of short-period pulsating variable stars with periods between 0.0125 and 0.25 days and amplitudes from milli-magnitude to tenths of a magnitude \citep{Breger2000,Casey2013}. They locate on either the pre-main sequence, or the main sequence, or the post-main sequence evolutionary stages, lying at the bottom of the classical Cepheid instability strip with spectral classes of A-F. The excitation mechanism of $\delta$ Scuti stars is the $\kappa$-mechanism which is the same as that of the Cepheid and the RR Lyrae stars \citep{Baker1962,Baker1965,Zhevakin1963}.

High-amplitude $\delta$ Scuti stars (hereafter HADS) are a subgroup of $\delta$ Scuti stars showing large amplitudes ($\Delta V \geq 0\fm3$) with single or double radial pulsation modes \citep[see, e.g.,\ ][]{Poretti2003,Niu2013,Niu2017}. Another subgroup, the SX Phoenicis variables, contains $\delta$ Scuti stars of Pop.\uppercase\expandafter{\romannumeral2}, being the old disk population showing large amplitudes like HADS. As the HADS generally rotate slowly with $v\sin i \leq 30$ km/s, \citet{Breger2000} indicated that the slow rotation seems to be a precondition for high amplitudes and possibly even for radial pulsations. \citet{Petersen1996} pointed out that the observed period ratios and positions on the H$-$R diagram of double-mode HADS are in agreement with the assumption that these variables are normal stars following standard evolution.

Theoretically, stellar evolution induces changes of periods in the pulsating stars. The secular observations of some variable stars show that the pulsation periods change in long time intervals. Whether the observed period change could be ascribed to the stellar evolution is an open question \citep{Rodriguez1995}. In principle, if the observation-determined period change rate of a variable falls into an interval derived from the model calculation, the observed period change is considered to be due to the stellar evolution effect.
For Classical Cepheids, this assumption helped Neilson estimate the mass loss and evolutionary stage of Polaris \citep{Neilson2012,Neilson2014}. In addition, \citet{Neilson2012_2} provided the first evidence that the enhanced mass loss must be a ubiquitous property of Classical Cepheids. For $\delta$ Scuti stars, \citet{Breger1998} predicted the values of $(1/P)(dP/dt)$ from $10^{-10}\ \mathrm{yr}^{-1}$ on the main sequence to $10^{-7}\ \mathrm{yr}^{-1}$ on the post-main sequence stages. With the same assumption, \citet{Yang2012} confirmed that the observed period change of the SX Phoenicis star XX Cyg could be successfully explained by the stellar evolution effect. Moreover, \citet{Niu2017} showed that the observed period change rate of the HADS AE UMa could not only be interpreted by the evolutionary effect, but also be regarded as an important observable that helps one perform successful asteroseismology on such kinds of stars.

VX\ Hydrae ($\alpha=09^h45^m46^s, \delta=-12^\circ00^\prime14^{\prime\prime}$) is a Pop.\uppercase\expandafter{\romannumeral1} HADS, discovered by \citet{Hoffmeister1931} and then observed by \citet{Lause1932} and \citet{Oosterhoff1938}. \citet{Fitch1966} derived the periods of the fundamental mode $P_{0}=0.223389$ days and of the first overtone $P_{1}=0.172720$ days by utilizing the time-series photometric observations. \citet{Breger1977} observed the star in $uvby$ and $\beta$ filters, measuring the variation ranges of the surface gravity $\log g \in [3.26,3.75]$ and of the effective temperature $T_\mathrm{eff} \in [6500,8050]$ K. The mean surface gravity $\overline {\log g} = 3.47$ and the mean effective temperature $\overline T_\mathrm{eff} = 7000$ K were also calculated. After re-analyzing the data of \citet{Breger1977}, \citet{McNamara1997} determined the mean effective temperature $\overline T_\mathrm{eff}=7100$ K, the metallicity ${\rm{[Fe/H]}}=0.05$, and the surface gravity $\log g \in [3.45,3.46]$. \citet{Templeton2009} monitored VX\ Hya in 2006 and 2007, determining the frequencies of the fundamental mode $f_{0}=4.4765\ \rm{c\ days^{-1}}$ and of the first overtone $f_{1}=5.7898\ \rm{c\ days^{-1}}$.
From the above frequency analysis results, one may note that (i) the frequency of fundamental mode of VX\ Hya (4.4765 $\rm{c\ days^{-1}}$) is smaller than that of most HADS \citep[see][Table 1]{McNamara2000}; (ii) the amplitude ratio of the fundamental and the first overtone modes of VX\ Hya $A_{f_{0}}/A_{f_{1}} \sim 1.3$ is obviously smaller than that of the other double-mode HADS (for AE UMa $\sim 5.5$, \citealt{Niu2017}; for RV Ari $\sim 3.2$, \citealt{Pocs2002}; etc.). These characteristics induced our interest to study VX\ Hya further based on our new observations.

The aim of this paper is to derive the stellar parameters and the evolutionary stage of VX\ Hya by means of asteroseismology. The paper is organized as follows. In Section 2, we report new photometric and spectroscopic observations and corresponding data reduction process. Section 3 presents the pulsation analysis and the period change rate calculation of VX\ Hya. In Section 4, we construct stellar evolution models and make pulsation frequency fitting. A brief discussion and the conclusions are presented in Sections 5 and 6, respectively.

\begin{table}
\caption{Journal of the Photometric Observations for VX\ Hya.}
\label{obs_jou}
\begin{center}
\begin{tabular}{lcccccccr}
\hline
\hline
Observatory & Longitude & Latitude & Telescope & Filter & Date & Nights & Frames\\
\hline
YAO & 102$\fdg$78E & 25$\fdg$03N & 101.6 cm & $V$ & 2015 Jan 25-Feb 4 & 10 & 2122\\
SPM & 117$\fdg$46W & 31$\fdg$04N & 84 cm & $V$ & 2015 Feb 2-6 & 5 & 1534\\
\hline
\end{tabular}
\end{center}
\end{table}

\section{observations and data reduction}

\subsection{Photometry}

\begin{table}\footnotesize
\caption{The Properties of VX\ Hya, the Comparison and the Check Star. }
\label{tab:com_che}
\begin{center}
\begin{tabular}{lccccc}
\hline
\hline
Star name                     & $\alpha$ (2000)   & $\delta$ (2000)               & $B$ (mag)           & $V$ (mag)        & $B-V$ (mag)     \\
\hline
Target = VX Hya               & $09^h45^m46\fs85$ & $-12^\circ00^\prime14\farcs3$ & $10.744\pm0.309$  & $10.485\pm0.205$   & $0.259\pm0.371$ \\
Comparison = TYC 5482-1225-1  & $09^h45^m50\fs92$ & $-11^\circ56^\prime46\farcs8$ & $12.656\pm0.009$  & $12.019\pm0.022$   & $0.637\pm0.024$ \\
Check = 2MASS 09454598-1201204 & $09^h45^m45\fs98$ & $-12^\circ01^\prime20\farcs5$ & $13.368\pm0.009$  & $12.882\pm0.021$  & $0.486\pm0.023$ \\
\hline
\multicolumn{6}{c}{Note. The $B$ and $V$ data were taken from AAVSO Photometric All Sky Survey (APASS) catalog \citep{APASS2016}.}
\end{tabular}
\end{center}
\end{table}

\begin{figure}
\centering
\includegraphics[width=0.5\textwidth,height=0.5\textwidth]{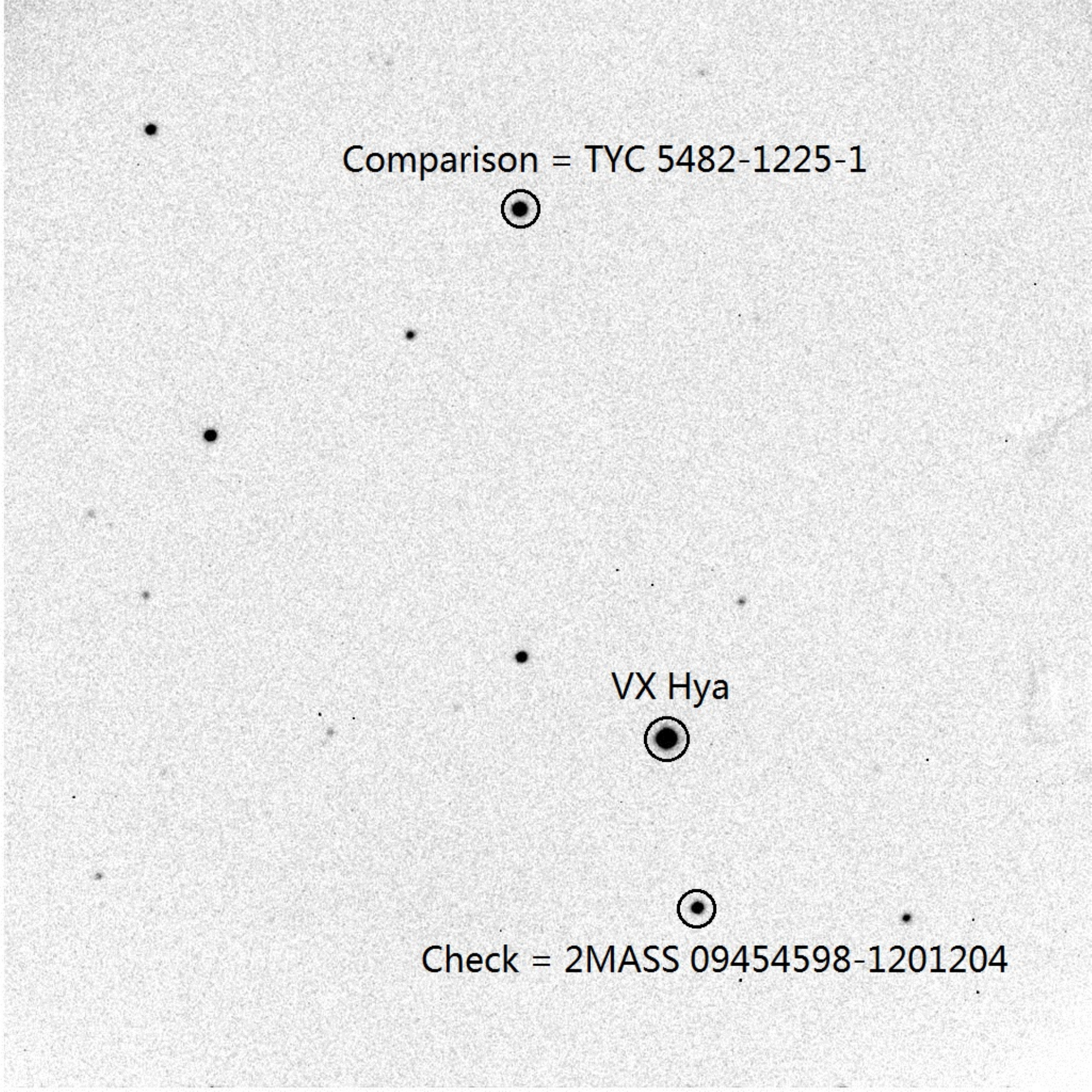}
\caption{A CCD image of VX\ Hya collected with 101T. The field of view is 7.3$^\prime\times$7.3$^\prime$. North is up and east is to the left. VX\ Hya, the comparison star, and the check star are marked.}
\label{CCDimage}
\end{figure}

VX Hya was observed in Johnson $V$ with the 101.6 cm telescope at Yunnan Astronomy Observatory (YAO) in China (hereafter 101T) and the 84 cm telescope at Observatorio Astron$\acute{o}$mico Nacional at Sierra San Pedro M$\acute{a}$rtir (OAN-SPM) in Mexico (hereafter 84T), from 2015 January to February. 101T and 84T were equipped with an Andor DW436 CCD camera and an E2V-4240 CCD camera, respectively. Both cameras had 2048$\times$2048 square pixel arrays with the resolution of 13.5 $\mu$m/pixel. The effective fields of view were $7.3^\prime\times7.3^\prime$ and $7.6^\prime\times7.6^\prime$ for 101T and 84T, respectively. Table \ref{obs_jou} lists the journal of the photometric observations. 2122 frames were collected over 10 nights with 101T, while 1534 frames were collected over 5 nights with 84T.

\begin{figure*}
\centering
\includegraphics[width=1.0\textwidth]{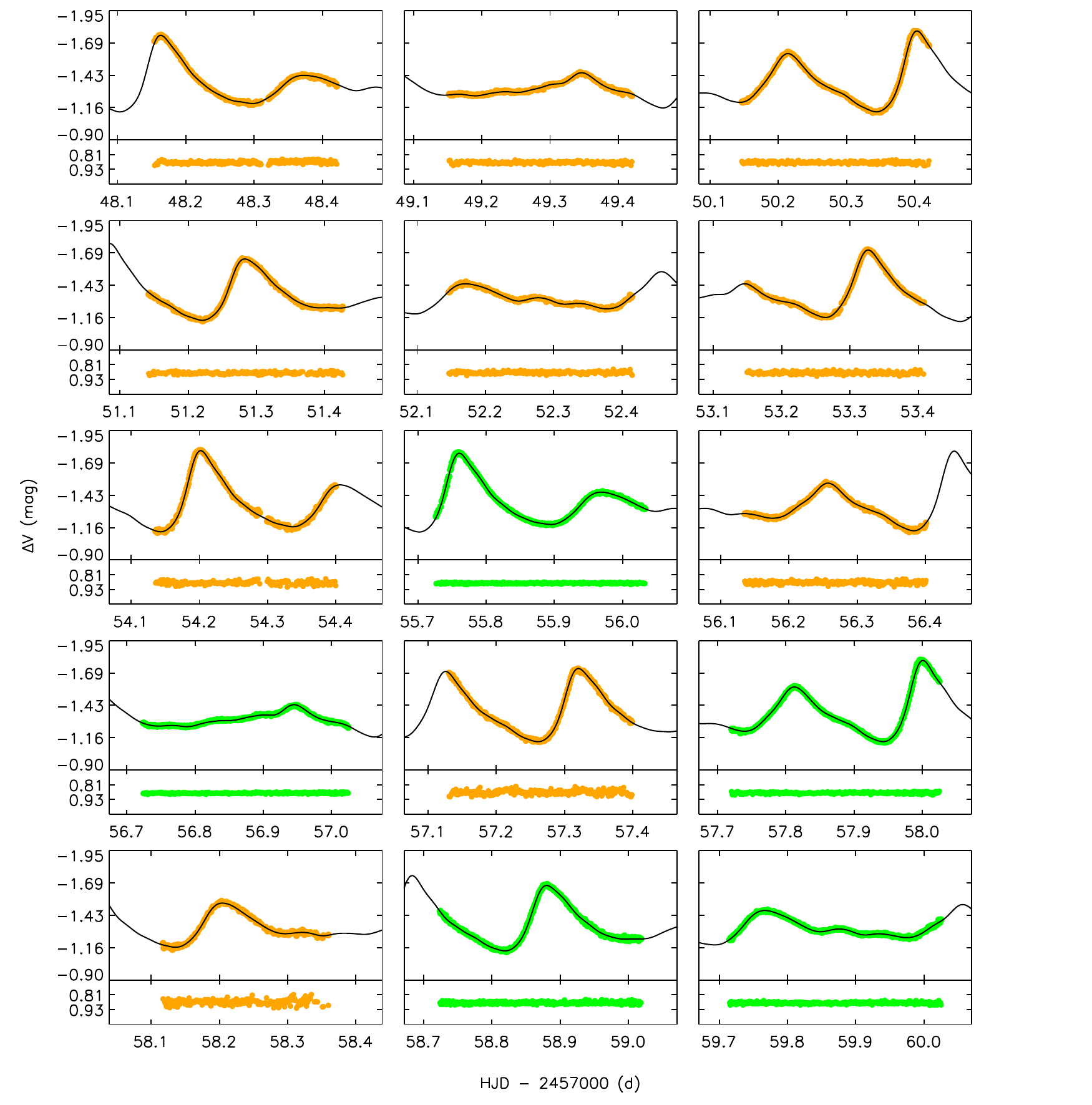}
\caption{Light curves of VX\ Hya in $V$ band in 2015. The top panels in each subfigure give the magnitude differences between VX\ Hya and the comparison star, while the bottom panels show the magnitude differences between the check star and the comparison star. The orange points come from the observations with 101T and the green ones from those with 84T. The solid curves represent the fitting with the 25-frequency solution listed in Table \ref{fre_solu}.}
\label{lighecurve}
\end{figure*}

Two stars on the same CCD image close to the target star VX\ Hya were used as the comparison star (TYC 5482-1225-1) and the check star (2MASS 09454598-1201204), respectively. Their brightnesses are similar to that of the target and stable during our observations (see Table \ref{tab:com_che} and Figure \ref{CCDimage}). Figure \ref{CCDimage} shows a CCD image taken with the 101T with the target star, the comparison and the check star marked. Table \ref{tab:com_che} lists their properties. The CCDRED routine of IRAF\footnote{Image Reduction and Analysis Facility is developed and distributed by the National Optical Astronomy Observatories, which is operated by the Association of Universities for Research in Astronomy under operative agreement with the National Science Foundation.} was used to subtract the bias and dark, and divide the flat for all the frames. After that, we performed aperture photometry for all images by utilizing the APPHOT routine of IRAF. Then, the magnitude differences between VX\ Hya and the comparison star were calculated together with those between the check star and the comparison star. The standard deviations of the differential magnitudes between the check star and the comparison star yield the estimations of the photometric precisions, with typical values of $0\fm004$ and $0\fm02$ in good and poor observational conditions, respectively. The zero-point differences among individual nights with typical values of 0.001-0.004 mag, which might be caused by the transparency instability of the atmosphere during the observation campaign, are compensated. Figure \ref{lighecurve} shows the light curves of VX\ Hya in $V$ band in 2015.

\subsection{Spectroscopy and Atmospheric Parameter Calculation}

\begin{deluxetable}{ll}
\tablecaption{Characteristics of the Spectroscopic Observations of VX\ Hya and Calculated Stellar Parameters. \label{tab:spec_para}}
\tablehead{
\colhead{Property} & \colhead{Value}
}
\startdata
Magnitudes ($V$)\tablenotemark{a}                         & 10.485             \\
Spectral Range (\AA)                                      & $3200 \sim 10000$   \\
$R$                                                       & 31,500             \\
S/N                                                       & 75                 \\
$\pi$ (mas)\tablenotemark{b}                              & $1.16\pm0.25$      \\
BC\tablenotemark{c}                                       & $-0.014$            \\
$A_V$\tablenotemark{d}                                    & $ 0.097$           \\
$\rm{[Fe/H]}$                                             & $-0.20\pm0.10$     \\
$T_\mathrm{eff}$ $_\mathrm{spec}$ \ (K)                 & $7193 \pm80$       \\
$T_\mathrm{eff}$ $_\mathrm{photometric}$\ (K)           & $7155 \pm96$       \\
$\rm{\log g}$ $_\mathrm{spec}\ \rm{(cm\,s^{-2})} $    & $ 3.50\pm0.20$     \\
$\rm{\log g}$ $_\mathrm{trigonometric}\ \rm{(cm\,s^{-2})}$    & $ 3.51\pm0.20$   \\
$\xi_\mathrm{t}\ \rm{(km\,s^{-1})}$                      & $ 2.90\pm0.20$      \\
$v\sin{i}\ \rm{(km\,s^{-1})}$                                & $ 6.6\pm0.2$    \\
$\rm{V_{rad}} $ $\rm{(km\,s^{-1})}$                        & $ -22.77$          \\
Mass/$M_\odot$                                            & $ 1.88\pm0.15$     \\
\enddata
\tablecomments{The $R$ and S/N indicate the spectrum resolution and S/N, respectively. $\pi$ is the parallax. $\xi_\mathrm{t}$ is the micro-turbulence velocity.}
\tablenotetext{a}{Magnitudes was taken from APASS catalog \citep{APASS2016}.}
\tablenotetext{b}{Parallax was taken from Gaia DR1 \citep{Gaia2016}.}
\tablenotetext{c}{Bolometric correction was derived from \citet{Alonso1999}.}
\tablenotetext{d}{Extinction in $V$ magnitude was taken from \citet{Schlafly2011}.}
\end{deluxetable}

\begin{figure}
 \centering
   \includegraphics[width=0.6\textwidth]{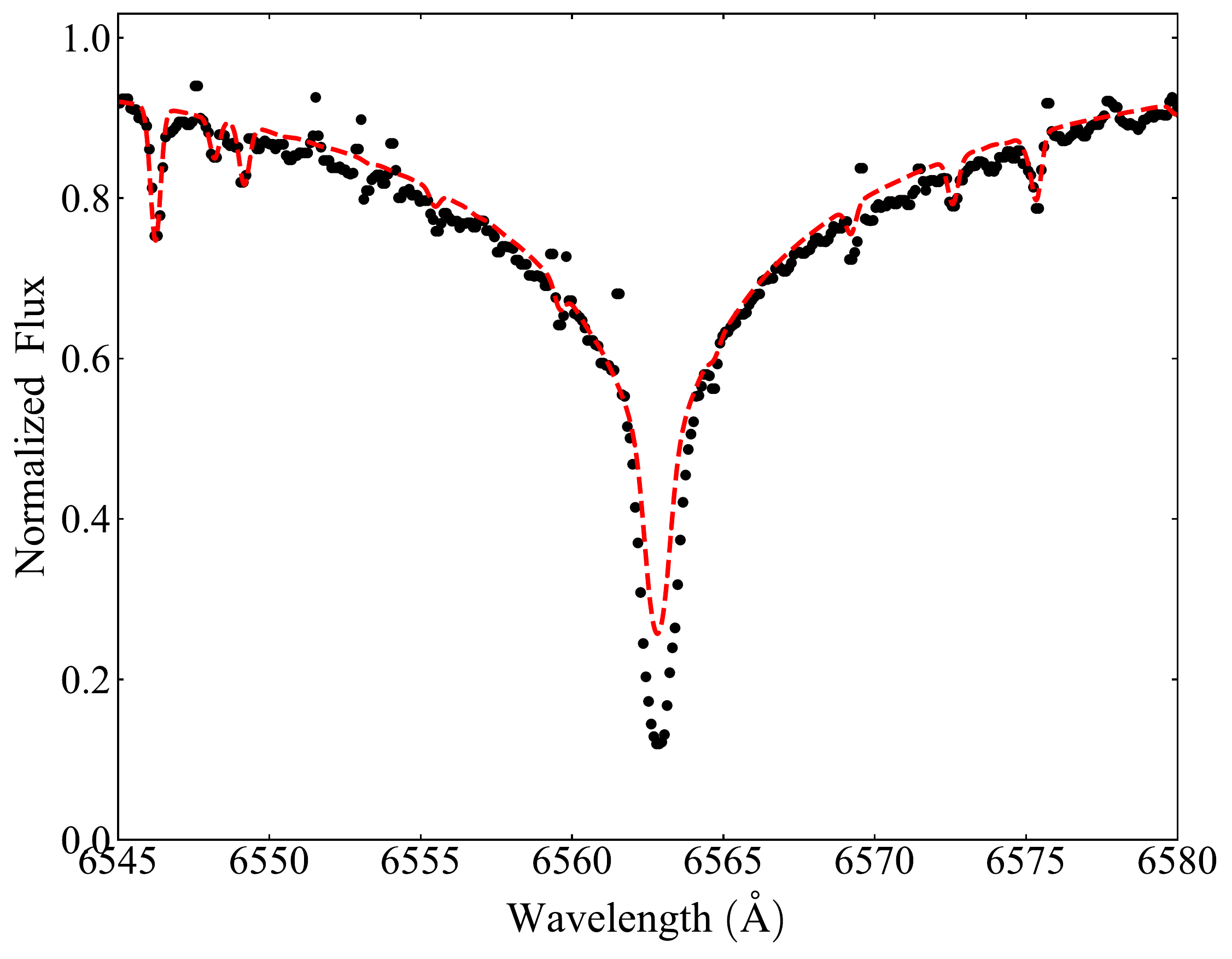}
      \caption{Spectral region covering the H$\alpha$ line profile of VX Hya. The observed spectrum is shown by the black dots, while the synthetic spectrum with the adopted atmospheric parameters is presented by the red dashed line.}
         \label{fig:Halpha}
   \end{figure}

\begin{figure}
 \centering
   \includegraphics[width=0.6\textwidth]{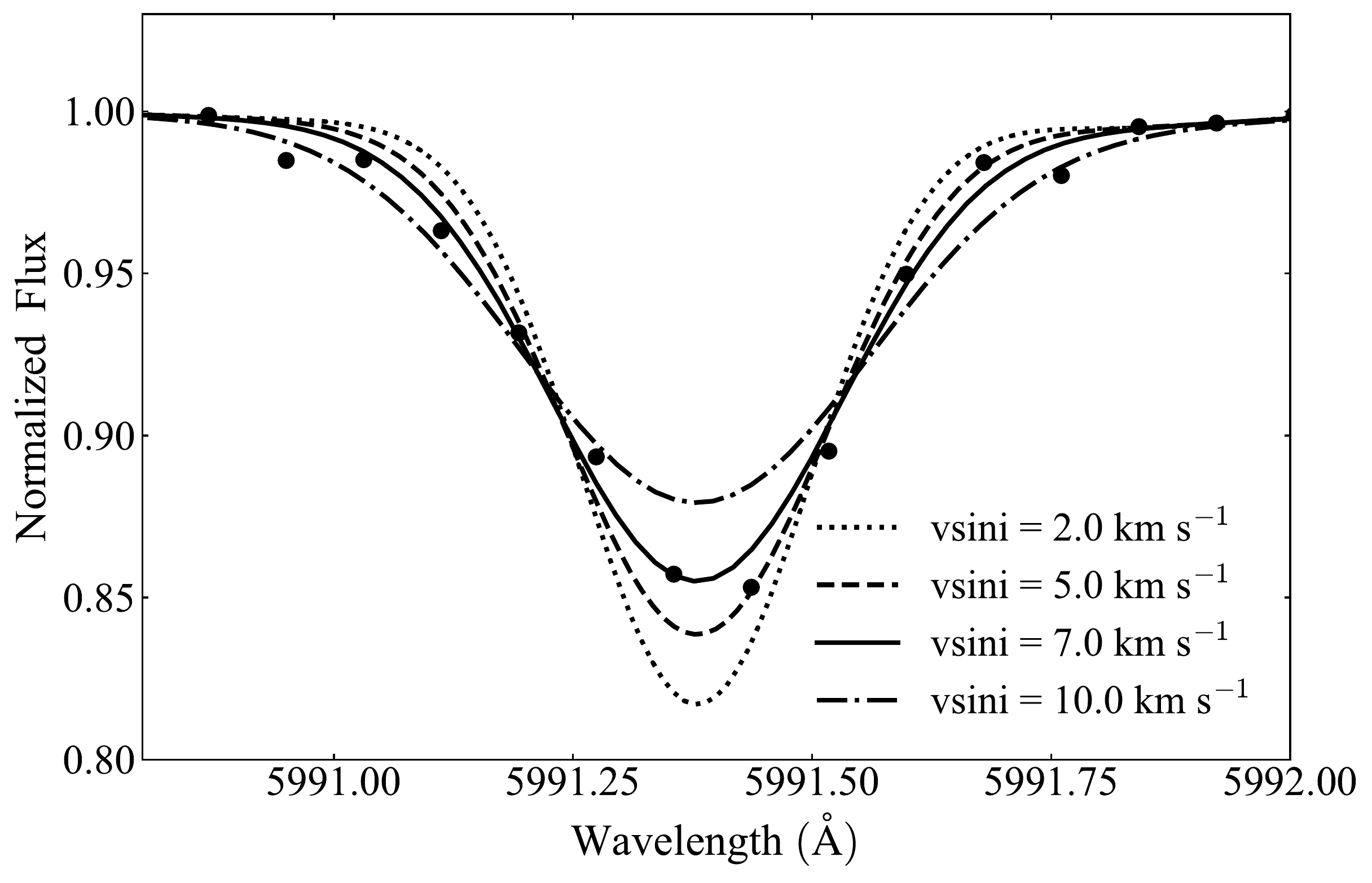}
      \caption{Spectral synthesis of Fe\,\uppercase\expandafter{\romannumeral1} line at 5991 \AA\ with individual values of $v\sin{i}$.}
         \label{fig:vsini}
   \end{figure}

 A high-resolution, high-signal-to-noise ratio (S/N) spectrum of VX\ Hya was obtained with the ARC \'{e}chelle Spectrograph (ARCES) mounted on the 3.5~m telescope located at Apache Point Observatory (APO) on 2015 December 08.\footnote{Based on observations obtained with the Apache Point Observatory 3.5-meter telescope, which is owned and operated by the Astrophysical Research Consortium.} The wavelength range of the instrument was 3200-10000 \AA\ with the resolution of $R\sim$ 31,500. The main characteristics of the observations are listed in Table \ref{tab:spec_para}.

Following the standard data extraction process, the raw data were reduced based on the improved standard automatic IDL program, which was initially designed for the fiber-coupled Cassegrain \'{e}chelle spectrograph (FOCES; \citealt{Pfeiffer1998}).

We employed the one-dimensional (1D) plane-parallel MAFAGS-OS atmospheric model to the spectral analysis, which was described in detail by \citet{Grupp2009}. In our analysis, the spectral synthesis method was applied to derive the abundances with Spectrum Investigation Utility (SIU). This code was based on the IDL program \citep{Reetz1991}. The spectroscopic method was adopted to determine the stellar parameters. The effective temperature was derived from the excitation equilibrium of Fe\,\uppercase\expandafter{\romannumeral1} lines with the excitation energy higher than 2.0 eV. The surface gravity was approached by requiring the same Fe abundances obtained from Fe\,\uppercase\expandafter{\romannumeral1} and Fe\,\uppercase\expandafter{\romannumeral2} lines. The micro-turbulence velocity $\xi_\mathrm{t}$ was determined when the Fe abundance derived from the individual Fe\,\uppercase\expandafter{\romannumeral1} lines does not depend on the equivalent widths. During our analysis, the nonlocal thermodynamic equilibrium (NLTE) effects on the Fe\,\uppercase\expandafter{\romannumeral1} lines had been considered.
We also derived the photometric $T_\mathrm{eff}$ from the relation between the intrinsic color $V-K$ and $T_\mathrm{eff}$ for the giant \citep{Alonso1999}, which is consistent with the spectroscopic one. As the H$\alpha$ line wing is sensitive to the $T_\mathrm{eff}$ variation, it is a good indicator to examine the effective temperature. Figure \ref{fig:Halpha} shows the comparison between the observed spectrum and the synthetic spectrum calculated with the adopted atmospheric parameters. The H$\alpha$ profile can be reproduced by the adopted $T_\mathrm{eff}$.

We note that Gaia DR1 \citep{Gaia2016} provided the parallax value of this object; thus, we also calculated the trigonometric $\log g$ with the equation:
\begin{equation}
\log g = \log {{M}\over {M_\mathrm\odot}}+ \log g_\mathrm\odot + 4\log {{T_\mathrm{eff}}\over {T_\mathrm{eff \odot}}} + 0.4(M_\mathrm{bol} - M_\mathrm{bol \odot}).
\end{equation}
Here, $M$ stands for the stellar mass, which was estimated with the Bayesian method \citep{daSilva2006} by using the PARSEC isochrones \citep{Bressan2012}. The method requests a set of spectro-photometric parameters (i.e., $T_\mathrm{eff}$, [Fe/H], $Vmag$, parallax) to estimate the stellar mass, age, surface gravity, radius, etc. By comparing with the theoretical isochrones, the probability distribution functions (PDF) of each stellar parameter were computed over the stellar isochrones. Then, the parameters could be determined when the PDFs present a single well-defined peak. The absolute bolometric magnitude $M_\mathrm{bol}$ is defined as
\begin{equation}
M_\mathrm{bol} = V_\mathrm{mag} + BC +5\log \pi + 5 - A_\mathrm{v}.
\end{equation}
where the parallax was taken from the Gaia DR1 \citep{Gaia2016}. The bolometric correction was derived from the empirical calibration of \citet{Alonso1999}, and the extinction in $V$ magnitude was estimated from the Galactic dust extinction \citep{Schlafly2011}. Within the uncertainties, the surface gravities derived from both methods are in a good agreement with each other.

The metallicity [Fe/H] was finally adopted until the interactive process converged, and the stellar parameters with the corresponding uncertainties are presented in Table \ref{tab:spec_para}. The isolated 29 Fe\,\uppercase\expandafter{\romannumeral1} and 8 Fe\,\uppercase\expandafter{\romannumeral2} lines had been adopted, which are listed in Table \ref{tab:atomic} in the Appendix, for the determination of the stellar parameters. The atomic line data were obtained from the Vienna Atomic Line Database (VALD; \citealt{Kupka1999}), and the oscillator strength (log $gf$) values were derived by fitting the solar spectrum.

In order to determine the projected rotational velocity ($v\sin{i}$), four relatively isolated iron lines at 5198, 5576, 5991, and 6065 \AA\ were adopted from the lines listed in Table \ref{tab:atomic} in the Appendix. When fitting the line profile, the broadenings caused by the instrumental broadening, the macroturbulence and $v\sin{i}$\ were included. The instrumental broadening was estimated from the Th-Ar lines with the Gaussian profile. The macroturbulent velocity was assumed to be 5 $\rm{km\,s^{-1}}$ \citep[for F subgiants]{Fekel1997}. We kept the instrumental broadening and macroturbulence fixed, and the $v\sin{i}$\ was derived until the best match between the synthetic and observed spectra was achieved. The synthetic spectra with the individual $v\sin{i}$\ are illustrated in Figure \ref{fig:vsini}.

\section{pulsation analysis}

\subsection{Frequency Analysis}

\setcounter{table}{4}
\begin{table}
\caption{Multi-frequency Solution of the Light Curves of VX\ Hya in $V$ Band in 2015.}
\label{fre_solu}
\centering
\begin{tabular}{lccccr}
\hline
\hline
 {No.}&{Identification}&{Frequency\ ($\rm{c\ days^{-1}}$)}&{Amplitude\ (mmag)}&{S/N}\\
\hline
F1   &$f_{0}$         & 4.4763(4)  &   143.2(2) & 349.1\\
F2   &$f_{1}$         & 5.7897(5)  &   110.0(2) & 303.3\\
F3   &$f_{0}+f_{1}$   &10.2659(3)  &    51.2(2) &  183.4\\
F4   &$2f_{0}$        & 8.952(1)   &    38.5(2) &  141.3\\
F5   &$f_{1}-f_{0}$   & 1.312(1)   &    34.2(2) &  104.5 \\
F6   &$2f_{1}$        &11.580(2)   &    21.8(2) &  64.8 \\
F7   &$f_{0}+2f_{1}$  &16.056(2)   &    15.4(2) &  57.9 \\
F8   &$2f_{0}-f_{1}$  & 3.164(3)   &    12.5(2) &  31.0 \\
F9   &$2f_{0}+f_{1}$  &14.742(3)   &    11.0(2) &  30.9 \\
F10  &$3f_{0}$        &13.429(3)   &     9.6(2) &  23.6 \\
F11  &$2f_{1}-f_{0}$  & 7.101(3)   &     8.8(3) &  30.3 \\
F12  &$3f_{1}$        &17.372(5)   &     5.7(2) &  24.7 \\
F13  &$f_{0}+3f_{1}$  &21.847(4)   &     5.6(2) &   26.1 \\
F14  &$3f_{0}+f_{1}$  &19.218(4)   &     5.1(2) &   19.5 \\
F15  &$2f_{0}+2f_{1}$ &20.531(5)   &     5.1(2) &   21.2 \\
F16  &$2f_{0}+3f_{1}$ &26.326(7)   &     3.3(2) &   11.3 \\
F17  &$3f_{0}-f_{1}$  & 7.635(8)   &     3.2(2) &   11.3 \\
F18  &$4f_{0}$        &17.900(8)   &     2.7(2) &   11.1 \\
F19  &$4f_{0}+f_{1}$  &23.692(9)   &     2.0(2) &   9.8 \\
F20  &$3f_{0}+2f_{1}$ &25.008(8)   &     2.4(2) &   11.5 \\
F21  &$f_{0}+4f_{1}$  &27.63(1)    &     2.2(2) &   6.6  \\
F22  &$4f_{1}$        &23.163(9)   &     2.2(2) &   9.9 \\
F23  &$3f_{0}+3f_{1}$ &30.79(1)    &     2.0(2) &   6.8 \\
F24  &$4f_{0}+2f_{1}$ &29.47(1)    &     1.8(2) &   7.4 \\
F25  &$2f_{0}+4f_{1}$ &32.10(1)    &     1.5(2) &   4.8 \\
\hline
\end{tabular}
\end{table}

\begin{figure*}
\centering
\includegraphics[width=1.0\textwidth]{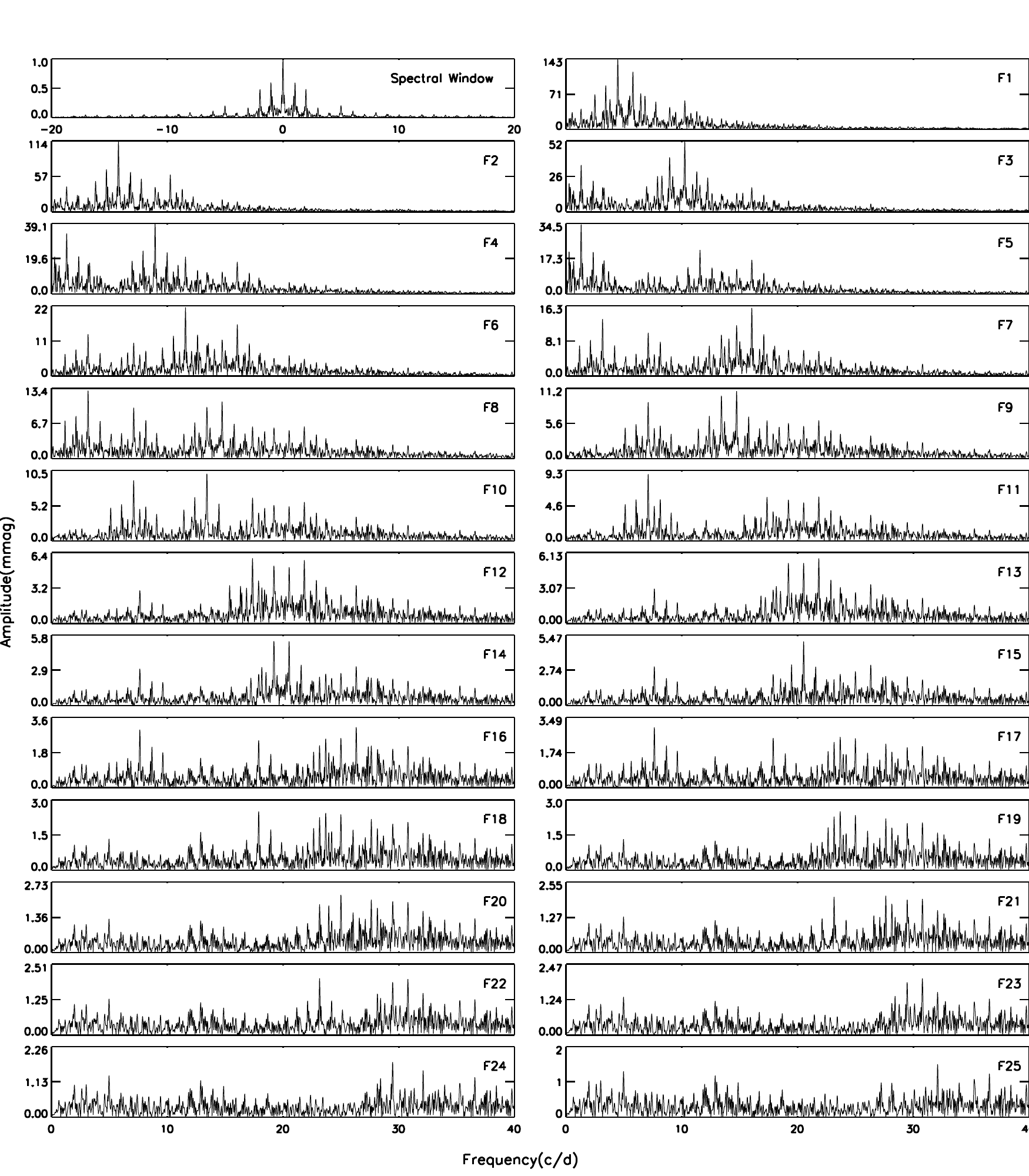}
\caption{Spectral window and Fourier amplitude spectra of the frequency pre-whitenning process for the light curves of VX\ Hya in V band in 2015.}
\label{pow_spec}
\end{figure*}

The software Period04 \citep{Lenz2005} was applied to make frequency analysis based on the Fourier transformations. Figure \ref{pow_spec} shows the spectral window and amplitude spectra of the frequency pre-whitenning process for the light curves of VX\ Hya in $V$ band in 2015.

The solution of 25 frequencies with S/Ns larger than 4.0 \citep{Breger1993} is listed in Table \ref{fre_solu}, including $f_{0}=4.4763\ \rm{c\ days^{-1}}$, $f_{1}=5.7897\ \rm{c\ days^{-1}}$ and 23 harmonics or linear combinations of $f_{0}$ and $f_{1}$. The errors of the frequencies and amplitudes in Table \ref{fre_solu} were estimated with the Monte Carlo simulations, which were based on the simulated light curves produced by an addition of the observed data and a Gaussian distribution random variable obeying $N(0,\sigma_{\rm{obs}})$ \footnote{Here, $\sigma_{obs}$ indicates the observation error.} (see Sect.~3 of \citet{Fu2013}). The solid curves in Figure \ref{lighecurve} show the fits with the multi-frequency solution.

The period ratio $P_{1}/P_{0} = f_{0}/f_{1} =0.7731$ derived from the observations in 2015, which agrees well with the result from \citet{Fitch1966}, indicates that $f_{0}$ is the frequency of the fundamental mode and $f_{1}$ the first overtone \citep{Petersen1996}.

\subsection{Period Change Rate}

\begin{table}
\caption{Journal of Different Groups of Light Curves.}
\label{tab:multiyear_obs}
\centering
\begin{tabular}{lccccc}
\hline
\hline
No.&Year   & Number of nights & Number of data points & Number of hours & References\\
\hline
1&1954     &      5           &     166              &  17.5      &\citet{Fitch1966}\\
2&1956     &      9           &     459              &  36.0      &\citet{Fitch1966}\\
3&1957     &      5           &     297              &  21.1      &\citet{Fitch1966}\\
4&1958     &      6           &     358              &  25.4      &\citet{Fitch1966}\\
5&2005     &      5           &      236              & 30.0      & AAVSO  \\
6&2006     &      19          &     3389              & 133.4      & AAVSO  \\
7&2008     &      7           &      345              &  48.2      & AAVSO  \\
8&2009     &      9           &      267              &  36.7      & AAVSO  \\
9&2010     &      17          &     3831              &  66.5      & AAVSO  \\
10&2011     &      5           &      313              & 20.9      & AAVSO  \\
11&2012     &      4           &      244              & 17.8      & AAVSO  \\
12&2013     &      6           &      325              & 21.1      & AAVSO  \\
13&2014    &      11          &      656              &  43.2      & AAVSO  \\
14&2015    &      15          &     3656              &  99.8      & this work \\
\hline
\multicolumn{6}{l}{Note. ``References'' indicates the souces of the data.}
\end{tabular}
\end{table}

\begin{deluxetable}{ccccc}
\tablecaption{Fourier Phases and Their Errors. }
\label{tab:Fourier_phase}
\tablehead{
\colhead{HJD$-$2430000} & \colhead{$\varphi_{0}$} & \colhead{$\sigma(\varphi_{0})$} & \colhead{$\varphi_{1}$} &   \colhead{$\sigma(\varphi_{1})$}
}
\startdata
   5158.30025     &         0.096   & 0.007      &   0.066   & 0.008\\
   5715.26675     &         0.148   & 0.006      &   0.112   & 0.007  \\
   6198.87175     &         0.132   & 0.005      &   0.141   & 0.008  \\
   6596.83505     &         0.111   & 0.004      &   0.050   & 0.005  \\
  23444.92019     &         0.176   & 0.005      &   0.060   & 0.006  \\
  23809.51435     &         0.189   & 0.002      &   0.094   & 0.002  \\
  24507.54468     &         0.166   & 0.005      &   0.134   & 0.006  \\
  24870.01302     &         0.134   & 0.006      &   0.106   & 0.007  \\
  25277.19806     &         0.140   & 0.001      &   0.118   & 0.002  \\
  25629.94859     &         0.133   & 0.006      &   0.137   & 0.006  \\
  25989.91636     &         0.127   & 0.007      &   0.119   & 0.007 \\
  26340.96562     &         0.143   & 0.005      &   0.090   & 0.006 \\
  26696.99443     &         0.128   & 0.004      &   0.097   & 0.005 \\
  27054.08968     &         0.116   & 0.001      &   0.096   & 0.002 \\
\enddata
\tablecomments{HJD is the middle heliocentric Julian date between the first and the last data point in each data group. $\varphi_{0}$ and $\varphi_{1}$ are the Fourier phases of $f_0$ and $f_1$, respectively. $\sigma(\varphi_{0})$ and $\sigma(\varphi_{1})$ are the corresponding uncertainties.}
\end{deluxetable}

\begin{figure*}
\centering
\gridline{\fig{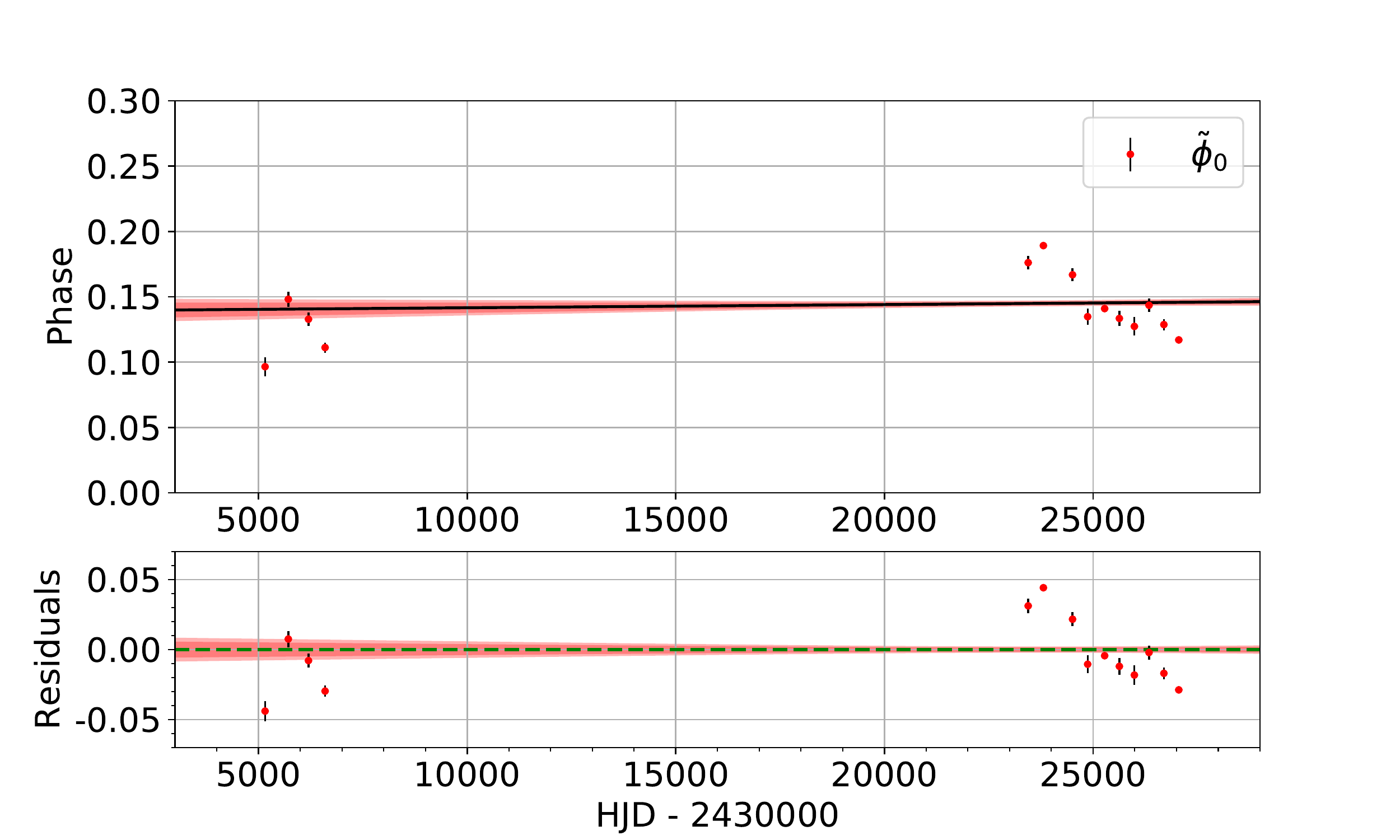}{0.5\textwidth}{(a)}
          \fig{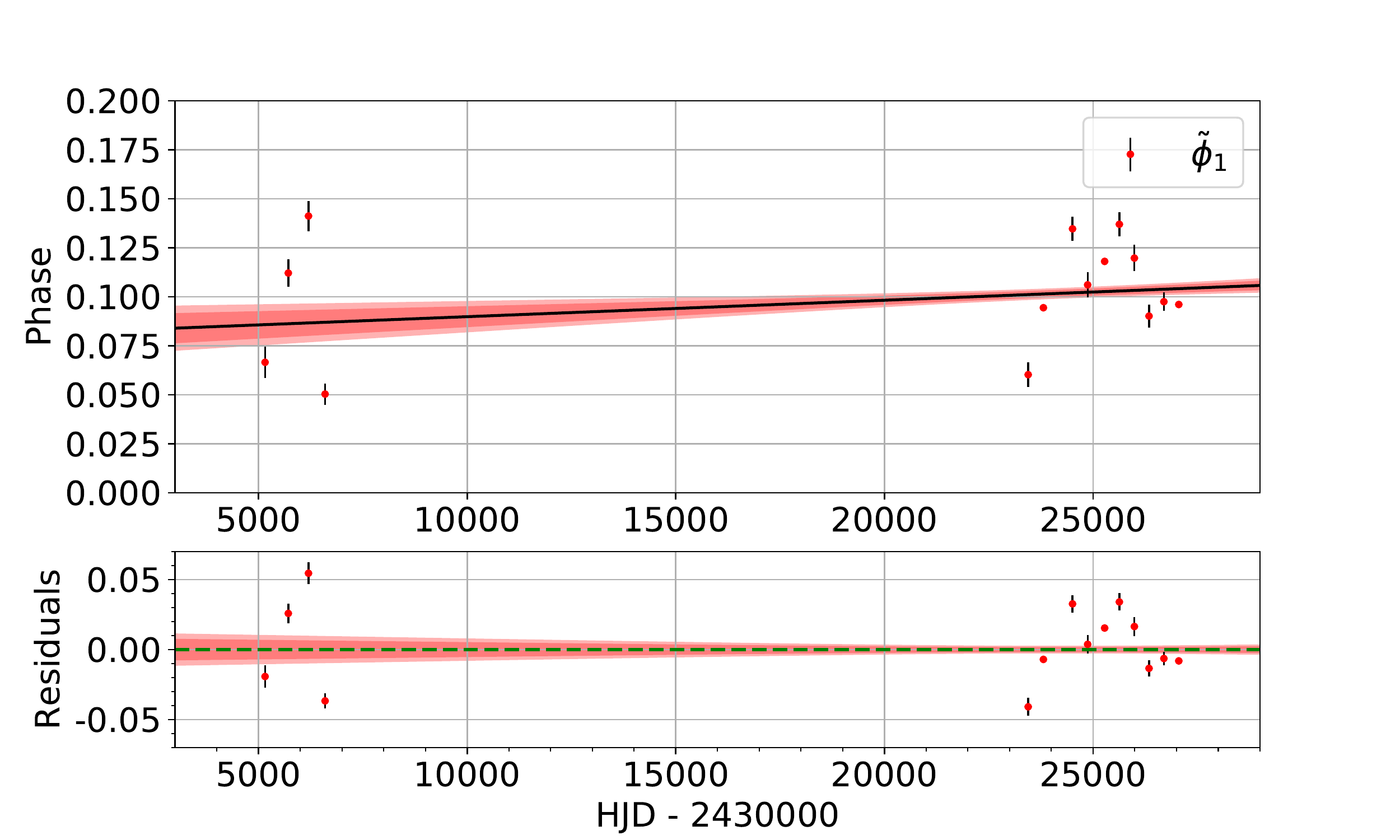}{0.5\textwidth}{(b)}}
\gridline{\fig{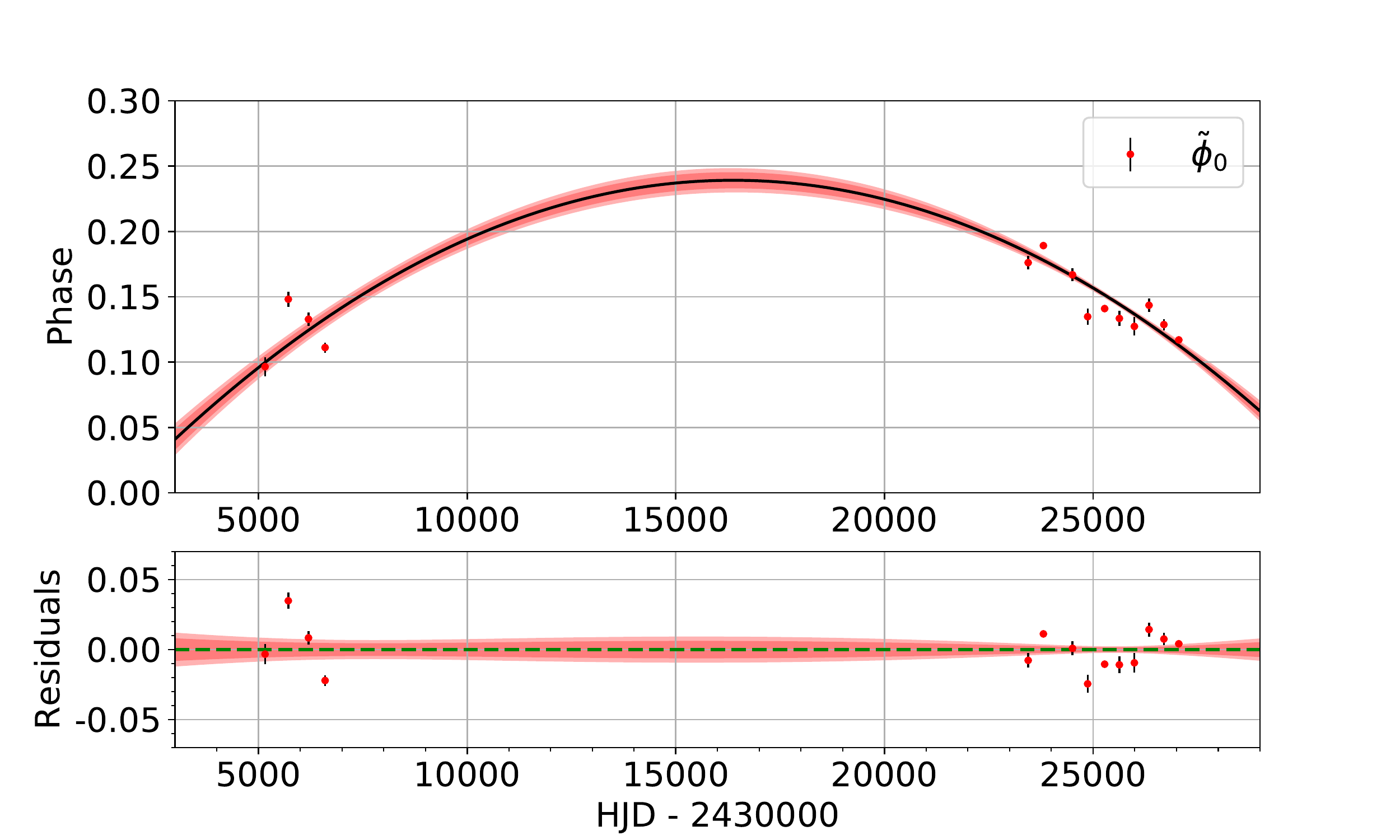}{0.5\textwidth}{(c)}
          \fig{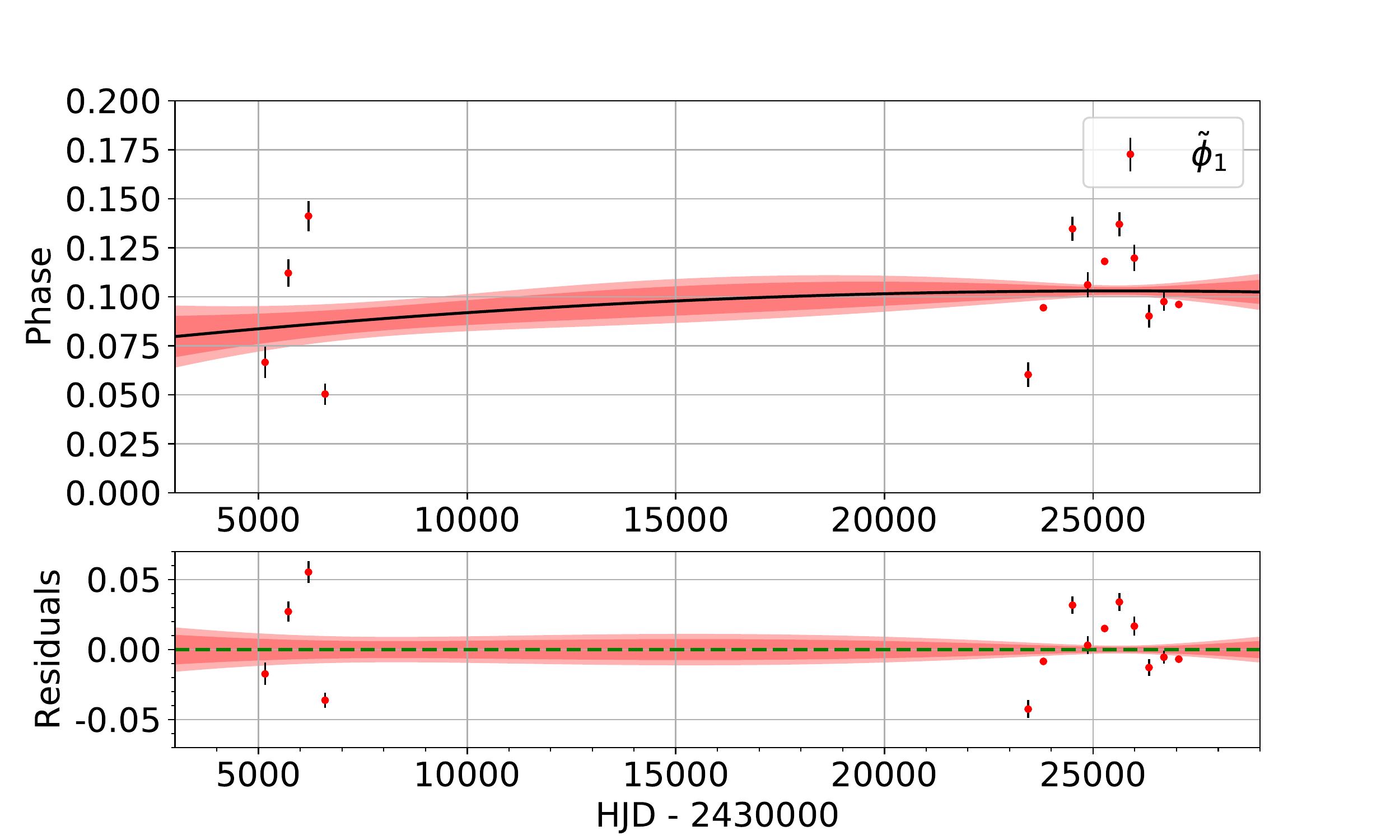}{0.5\textwidth}{(d)}}
\caption{The top panels of subfigure (a) and (c) show the linear and the quadratic fitting results of $\tilde{\phi}_{0}$, respectively. The top panels of (b) and (d) show the linear and the quadratic fitting results of $\tilde{\phi}_{1}$, respectively. The bottom panels of each subfigure show the corresponding fitting residuals. The 2$\sigma$ (deep red) and 3$\sigma$ (light red) bounds are also shown in the panels.}
\label{fig:phi_best_fit}
\end{figure*}

\begin{deluxetable}{ccc}
\caption{Linear and Quadratic Fitting Results for $\tilde{\phi}_{0}$ and $\tilde{\phi}_{1}$. }
\label{tab:params_fit}
\tablehead{
\colhead{Parameters} & \colhead{Fitting Results} & \colhead{$\chi^{2}/\rm{dof}$} 
}
\startdata
$a_{0}$& $-5(\pm 5) \times 10^{-2}$ & 20.86 \\
$b_{0}$& $3.6(\pm 0.2) \times 10^{-5}$ & \\
$c_{0}$& $-1.11(\pm0.06) \times 10^{-9}$ & \\
\hline
$d_{0}$& $1.4(\pm0.5) \times 10^{-1}$ & 116.23 \\
$e_{0}$& $0.6(\pm3.6) \times 10^{-6}$ & \\
\hline
$a_{1}$& $8(\pm8) \times 10^{-2}$ & 29.95 \\
$b_{1}$& $2(\pm2) \times 10^{-6}$ & \\
$c_{1}$& $-4(\pm7) \times 10^{-11}$ & \\
\hline
$d_{1}$& $9(\pm8) \times 10^{-2}$ & 27.57\\
$e_{1}$& $0.7(\pm1.8) \times 10^{-6}$ & \\
\enddata
\tablecomments{$a_i, b_i, c_i$ are the quadratic fitting parameters and $d_i, e_i$ are the linear fitting parameters of $\tilde{\phi}_{i}$.}
\end{deluxetable}

Considering the similar amplitudes of $f_{0}$ and $f_{1}$ from the Fourier analysis (see Table \ref{fre_solu}), the traditional O$-$C method (see, e.g., \citet{Yang2012,Niu2017}) could not be used to determine the period change of $f_{0}$ which is effective only when the amplitude of the dominant frequency is much larger than that of the second strongest frequency. Consequently, we employed the Fourier-phase diagram method developed by \citet{Paparo1998} to calculate the period change rate of VX\ Hya. This method can help study not only the period change of $f_{0}$, but also that of $f_{1}$. \citet{Pocs2002} analyzed the period change of the double-mode HADS star RV Ari with this method and obtained a result that was consistent with that from the O$-$C method.

The Fourier decomposition can be presented by the formula,
\begin{equation}
\label{eq:Fourier_de}
m = m_{0} + \sum_{i} A_{i} \sin \left[ 2 \pi (f_{i} t + \phi_{i}) \right] = m_{0} + \sum_{i} A_{i} \sin \left[ 2 \pi \Phi_{i} \right] ~,
\end{equation}
where $A_{i}$ is the amplitude, $f_{i}$ the frequency, and $\phi_{i}$ the corresponding initial phase. Here, we define $\Phi_{i} = f_{i}t + \phi_{i}$. If one wants to investigate the linear variation of $f_{i}$, one should add a term in $\Phi_{i}$,
\begin{equation}
\label{eq:Phi}
\Phi_{i} = f_{i} t + \frac{1}{2} \dot{f}_{i} t^{2} + \phi_{i} ~.
\end{equation}

 With the Fourier-phase diagram method, one may fix the amplitudes and frequencies in Eq. (\ref{eq:Fourier_de}) and use the variations in $\phi_{i}$ to reflect the change of $f_{i}$. In that case, one has,
\begin{equation}
\label{eq:Phi_tidal}
\Phi_{i} = \hat{f}_{i} t  + \tilde{\phi}_{i} = \hat{f}_{i} t  + a_{i} + b_{i} t + c_{i} t^{2} ~.
\end{equation}
Here, $\hat{ }$ indicates fixed values, and $\tilde{\phi}_{i}$ the variation term in the method, which is represented by a quadratic polynomial ($\tilde{\phi}_{i} = a_{i} + b_{i} t + c_{i} t^{2}$).

Comparing Eq. (\ref{eq:Phi}) and (\ref{eq:Phi_tidal}), one gets the meanings of $a_{i}$, $b_{i}$, and $c_{i}$: $a_{i}$ represents a constant value of phase; $b_{i}$ a correction on the value of $\hat{f}_{i}$, while $\hat{f}_{i} + b_{i}$ is the corrected value of the frequency after fitting; $2 c_{i}$ the value of $\dot{f}_{i}$ in Eq. (\ref{eq:Phi}). Moreover, the standard deviation of $b_{i}$ and $2 c_{i}$ can be regarded as the standard deviation of $f_{i}$ and $\dot{f}_{i}$.

In order to apply this method, we collected light curves from \citet{Fitch1966}, the American Association of Variable Star Observers (AAVSO), and our observations,\footnote{All of these data have been observed in V band.} and divided them into 14 groups (see Table \ref{tab:multiyear_obs}). We used Period04 to extract the first two frequencies ($f_{0}$ and $f_{1}$), which have the largest amplitudes for the whole data sets. Then, for each group, we fixed the amplitudes and frequencies, and let the phases be a free parameter to fit the light curves. The nonlinear least-square method was used to calculate the best-fit values of the phases.\footnote{The errors of the phases were estimated by Monte Carlo simulations, similar to the method used in the frequency analysis.} The obtained phases ($\tilde{\phi}_{0}$ and $\tilde{\phi}_{1}$) and their errors for each group are listed in Table \ref{tab:Fourier_phase}.

For $\tilde{\phi}_{0}$ and $\tilde{\phi}_{1}$, we considered both the linear fitting ($\tilde{\phi}_{i} = d_{i} + e_{i} t$) and the quadratic fitting ($\tilde{\phi}_{i} = a_{i} + b_{i} t + c_{i} t^{2}$). The Markov Chain Monte Carlo method (MCMC; see \citealt{Sharma2017} for review) was used to determine the best-fit values of the parameters and their errors. The best-fitting results and the corresponding residuals of $\tilde{\phi}_{0}$ and $\tilde{\phi}_{1}$ are shown in Figure \ref{fig:phi_best_fit}. The mean values and the standard deviations of the parameters are listed in Table \ref{tab:params_fit}, where $a_i, b_i, c_i$ are the quadratic fitting parameters and $d_i, e_i$ the linear fitting parameters of $\tilde{\phi}_{i}$. With the values of the parameters $a_i, b_i, c_i$ in Table \ref{tab:params_fit}, we derived $f_{0} = 4.476478\pm0.000002\ \rm{c\ days^{-1}}$, $\dot{f}_{0} = -2.22 (\pm 0.12) \times 10^{-9}\ \rm{c\ days^{-2}}$, $f_{1} = 5.789769\pm0.000002\ \rm{c\ days^{-1}}$, and $\dot{f}_{1} = -8 (\pm 14) \times 10^{-11}\ \rm{c\ days^{-2}}$. Hence, the period change rates of $f_{0}$ and $f_{1}$ can be derived as follows,
\begin{equation}
\frac{1}{P_{0}}\frac{{\rm{d}}P_{0}}{{\rm{d}}t}=-\frac{1}{f_{0}}\frac{{\rm{d}}f_{0}}{{\rm{d}}t}=(1.81\pm0.09)\times10^{-7} \ \mathrm{yr^{-1}},
\end{equation}
\begin{equation}
\frac{1}{P_{1}}\frac{{\rm{d}}P_{1}}{{\rm{d}}t}=-\frac{1}{f_{1}}\frac{{\rm{d}}f_{1}}{{\rm{d}}t}=(5.05\pm8.83)\times10^{-9} \ \mathrm{yr^{-1}}.
\end{equation}

Comparing the $\chi^{2}/\rm{dof}$\footnote{Here, $\chi^{2} = \frac{({\bm O}_{th} - {\bm O}_{obs})^{2}}{\sigma_{obs}^{2}}$, where ${\bm O}_{th}$ and ${\bm O}_{obs}$ are the theoretically calculated and the observed values of an observable, respectively; $\sigma_{\rm{obs}}$ is the observed error. dof implies the degree of freedom, which is defined as the number of data points minus the number of free parameters in the fitting.} between the linear and quadratic fitting results, one can conclude that (i) for the fundamental mode, a quadratic fitting is necessary and a reliable value of $\dot{f}_{0}$ can be obtained; (ii) for the first overtone mode, it is not absolutely necessary to employ a quadratic fitting. As a result, we used only $\dot{f}_{0}$ as an effective constraint in our substantial stellar model calculation.

\section{constraints from theoretical models}

We used Modules for Experiments in Stellar Astrophysics (MESA \citealt{Paxton2011, Paxton2013, Paxton2015}), a suite of open source for computation in stellar astrophysics, to construct the theoretical models. The stellar evolution module MESA star combines a number of numerical and physical modules for simulations of stellar evolution. The $\rho-T$ tables are based on the updated OPAL EOS tables in 2015 \citep{Rogers2002} and extended to lower temperatures and densities by the SCVH tables \citep{Saumon1995}. HELM \citep{Timmes2000} and PC \citep{Potekhin2010} tables are used at the outside region of OPAL and SCVH. MESA opacity tables are composed of OPAL Type 1 and 2 tables \citep{Iglesias1993, Iglesias1996} for the high temperature region, tables of \citet{Ferguson2005} for the low temperature region, and tables from OP \citep{Seaton2005} as the table format is identical. The standard mixing-length theory of convection of \citet*[chap.14]{Cox1968} and the modified MLT of \citet{Henyey1965} are used in MESA. The overshooting mixing diffusion coefficient presented by \citet{Herwig2000} is adopted in MESA star. The adiabatic pulsation code ADIPLS \citep{Christensen-Dalsgaard2008} in MESA enables pulsation frequencies to be calculated.

\subsection{Pulsation Parameters}

\begin{table*}
\centering
\caption{Pulsation Parameters of VX\ Hya.}
\label{obs_para}
\begin{tabular}{ccc}
\hline
\hline
$f_{0}$ ($\rm{c\ days^{-1}}$) & $f_{1}$ ($\rm{c\ days^{-1}}$) & $(1/P_{0})(dP_{0}/dt)$ $(\rm{yr^{-1}})$  \\
\hline
$4.4763 \pm 0.0004$ & $5.7897 \pm 0.0005$ & $1.81 \pm 0.09~(10^{-7})$  \\
\hline
\end{tabular}
\end{table*}

The frequencies of the fundamental mode $f_0$ and of the first overtone $f_1$ determined from the observations in 2015 were utilized to restrict the theoretical models. In addition, as the value of period change rate of the fundamental mode $f_0$ is within the predicted values of $\delta$ Scuti stars \citep{Breger1998}, we assumed that the period change rate of VX\ Hya calculated in this paper is due to the stellar evolutionary effect, which was applied to help constrain the models. These values were utilized as pulsation parameters of VX\ Hya, as listed in Table \ref{obs_para}.

\subsection{Setup of the Model Calculation}

As pointed out by \citet{Poretti2005}, the period ratios $P_{1}/P_{0}$ of long-period HADS stars higher than 0.770 could only exist for the models with the mass $M$ larger than 2.00$M_{\odot}$, while one notes that $P_{1}/P_{0}=0.7731\pm0.0001$ for VX\ Hya according to Table~\ref{obs_para}. On the other hand, \citet{Breger1977} indicated that VX\ Hya should be a normal high-mass $\delta$ Scuti star. Conservatively, we constructed a series of stellar models with the mass range from 1.80 $M_{\odot}$ to 2.80 $M_{\odot}$ with the step of 0.01 $M_{\odot}$. The heavy-element mass fraction $Z$ was determined as 0.010 from the metallicity value $\mathrm{[Fe/H]}=-0.2$ dex as listed in Table~\ref{tab:spec_para}. The initial hydrogen abundance $X$ was set as 0.7 as usual. The value of the mixing-length parameter was taken as $\alpha_{\rm{{MLT}}}=1.77$ which has a slight effect on the models (c.f. \citet{Breger2000, Yang2012, Niu2017}). As VX\ Hya is a slow rotator with $v\sin{i}\sim6.6\ \rm{km~s^{-1}}$ (see Table \ref{tab:spec_para}), the effects of rotation were not considered in our calculations \citep{Breger2000, Petersen1998}. Each model of the evolution sequence started from the ZAMS to the post-main-sequence stage by specifying the mass $M$ and the initial chemical composition ($X$,$Y$,$Z$). From the stage of the main sequence to the post-main sequence, the ADIPLS program was invoked to calculate the pulsation frequencies of each evolutionary step. In our calculation, $f_{0}$ and $f_{1}$ were derived with the quantum numbers of ($\ell=0, n=1$) and ($\ell=0, n=2$), respectively.

\subsection{Parameter Fitting}

\begin{figure*}
\centering
\gridline{\fig{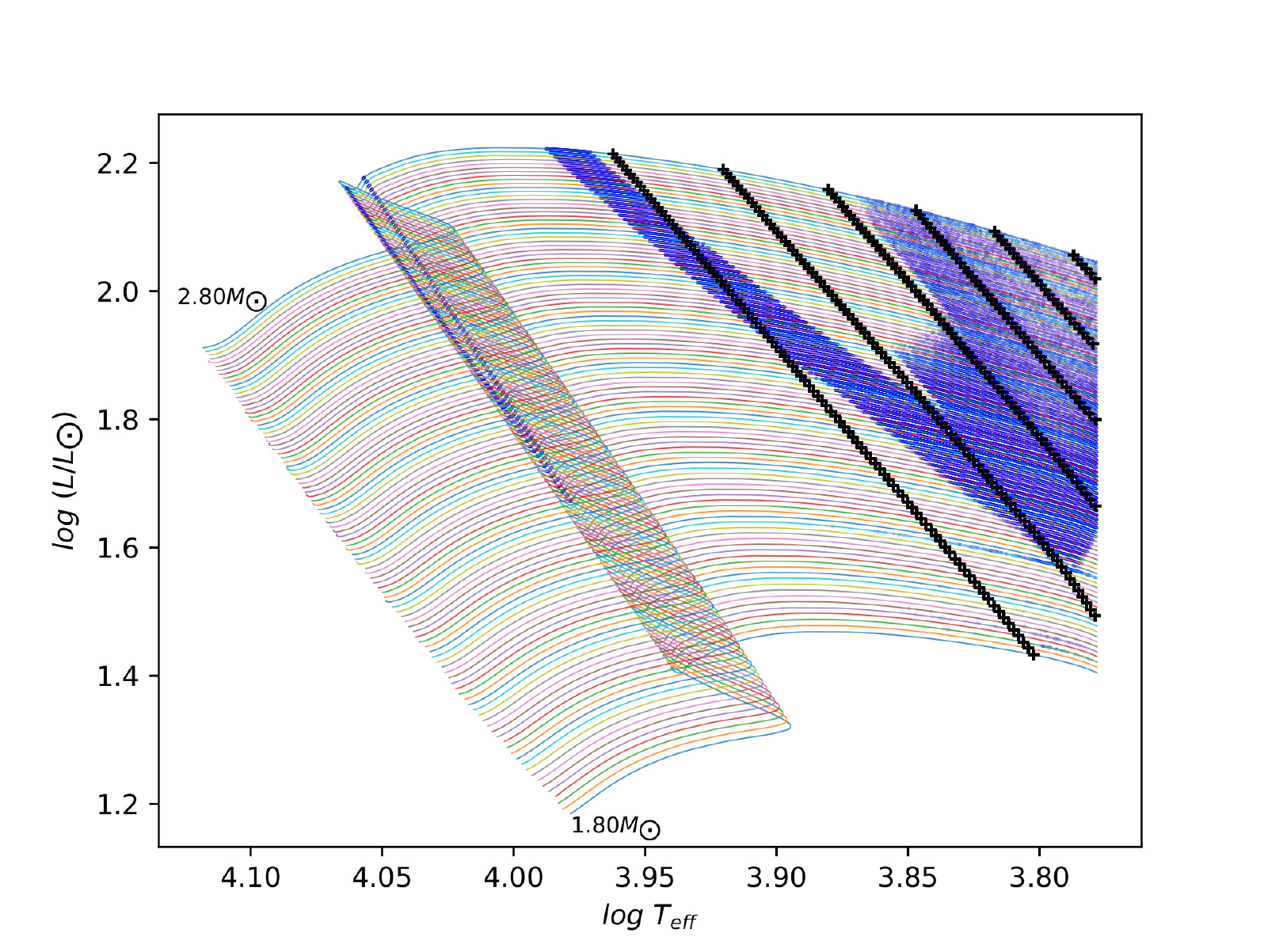}{0.48\textwidth}{(a)}
          \fig{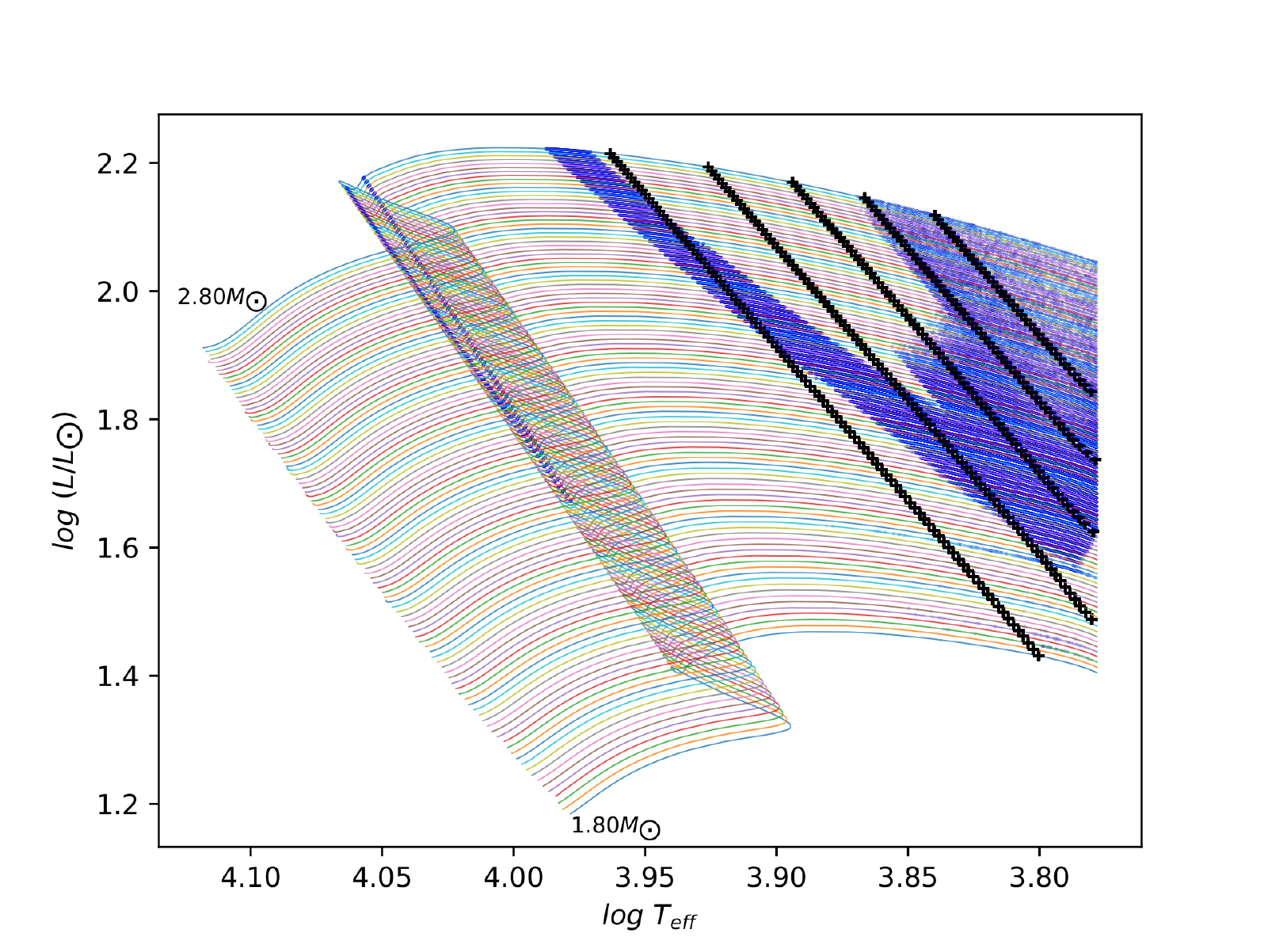}{0.48\textwidth}{(b)}}
\gridline{\fig{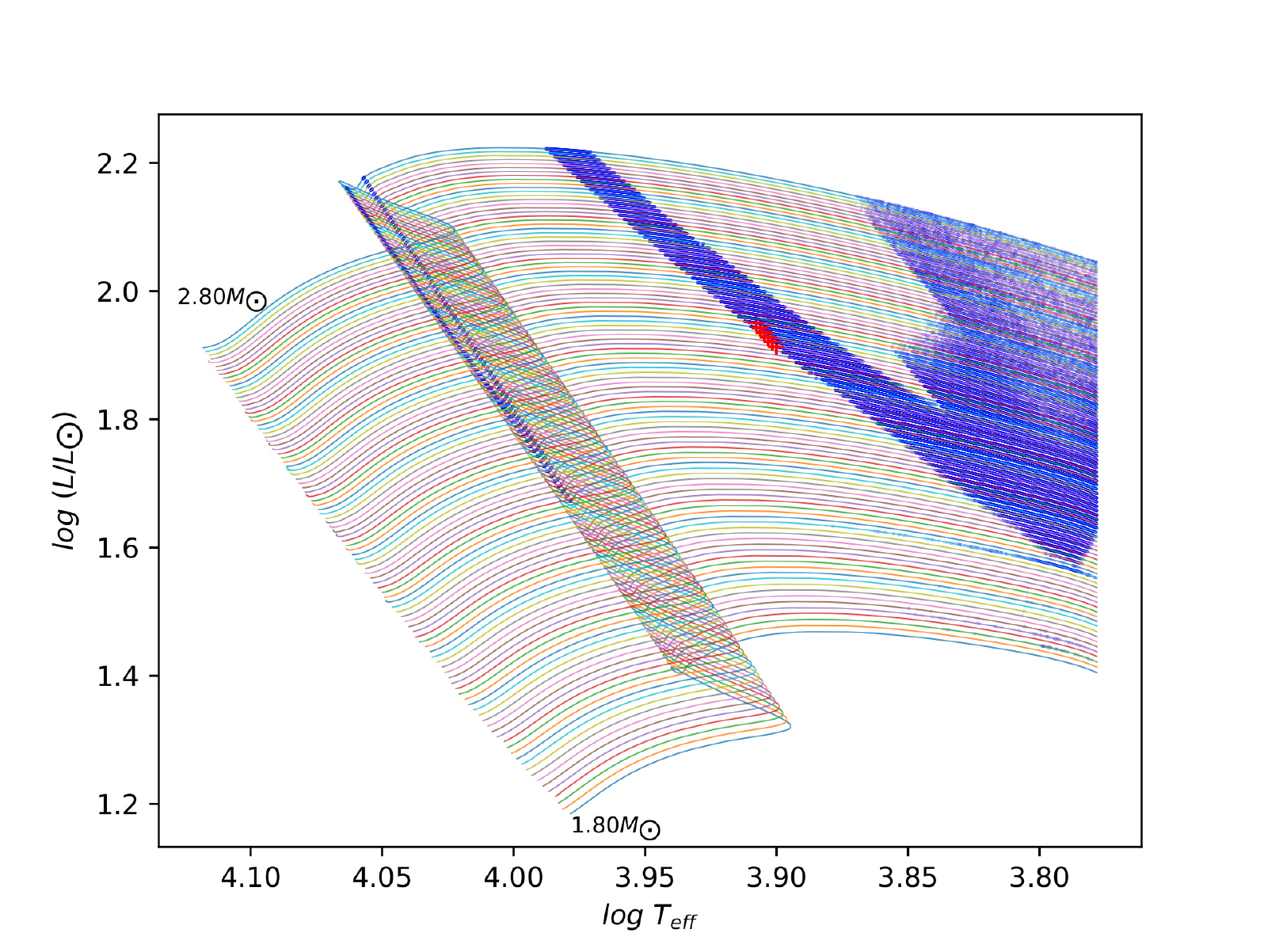}{0.65\textwidth}{(c)}}
\caption{The evolutionary tracks from the main sequence to the end of the post-main sequence in the mass range of 1.8-2.8 $M_{\odot}$ with the step of $0.01 M_{\odot}$. In all three subfigures, the blue regions on the tracks indicate the models for which the calculated $(1/P_{0})(dP_{0}/dt)$ fit the observation-determined $(1/P_{0})(dP_{0}/dt)$. In subfigure (a), the black lines mark the models for which the calculated $f_0$ fit the observation-determined $f_0$; in subfigure (b), the black lines mark the models for which the calculated $f_1$ fit the observation-determined $f_1$; in subfigure (c), the red lines mark the models for which the calculated $f_0$ and $f_1$ fit simultaneously the observation-determined $f_0$ and $f_1$. On these tracks, the observation-determined $f_{0}$, $f_{1}$ and $(1/P_{0})(dP_{0}/dt)$ are marked within $3\sigma$.}
\label{hr}
\end{figure*}

  The evolutionary tracks from the main sequence to the end of the post-main sequence in the mass range of 1.8-2.8 $M_{\odot}$ are shown in Figure \ref{hr}. In all the three subfigures, the blue regions on the tracks indicate the models for which the calculated $(1/P_{0})(dP_{0}/dt)$ fit the observation-determined $(1/P_{0})(dP_{0}/dt)$. In subfigure (a), the black lines mark the models for which the calculated $f_0$ fit the observation-determined $f_0$; in subfigure (b), the black lines mark the models for which the calculated $f_1$ fit the observation-determined $f_1$; in subfigure (c), the red lines mark the models for which the calculated $f_0$ and $f_1$ fit simultaneously the observation-determined $f_0$ and $f_1$. On these tracks, the observation-determined $f_{0}$, $f_{1}$ and $(1/P_{0})(dP_{0}/dt)$ are marked within $3\sigma$\footnote{Here, $\sigma$ values are the standard deviations of the quantities determined from observations.}. More details about the fitted models can be found in Table \ref{tab:fitting_results} in the Appendix\footnote{Of course, the number of fitted models depends on the evolutionary steps and the initial mass steps that we have chosen. However, we think at present configurations, it is sufficient for us to get some valuable results.}. The physical parameters of VX\ Hya derived from the fitted models are listed in Table \ref{tab:best_par}.

\begin{figure*}
\centering
\includegraphics[width=0.7\textwidth]{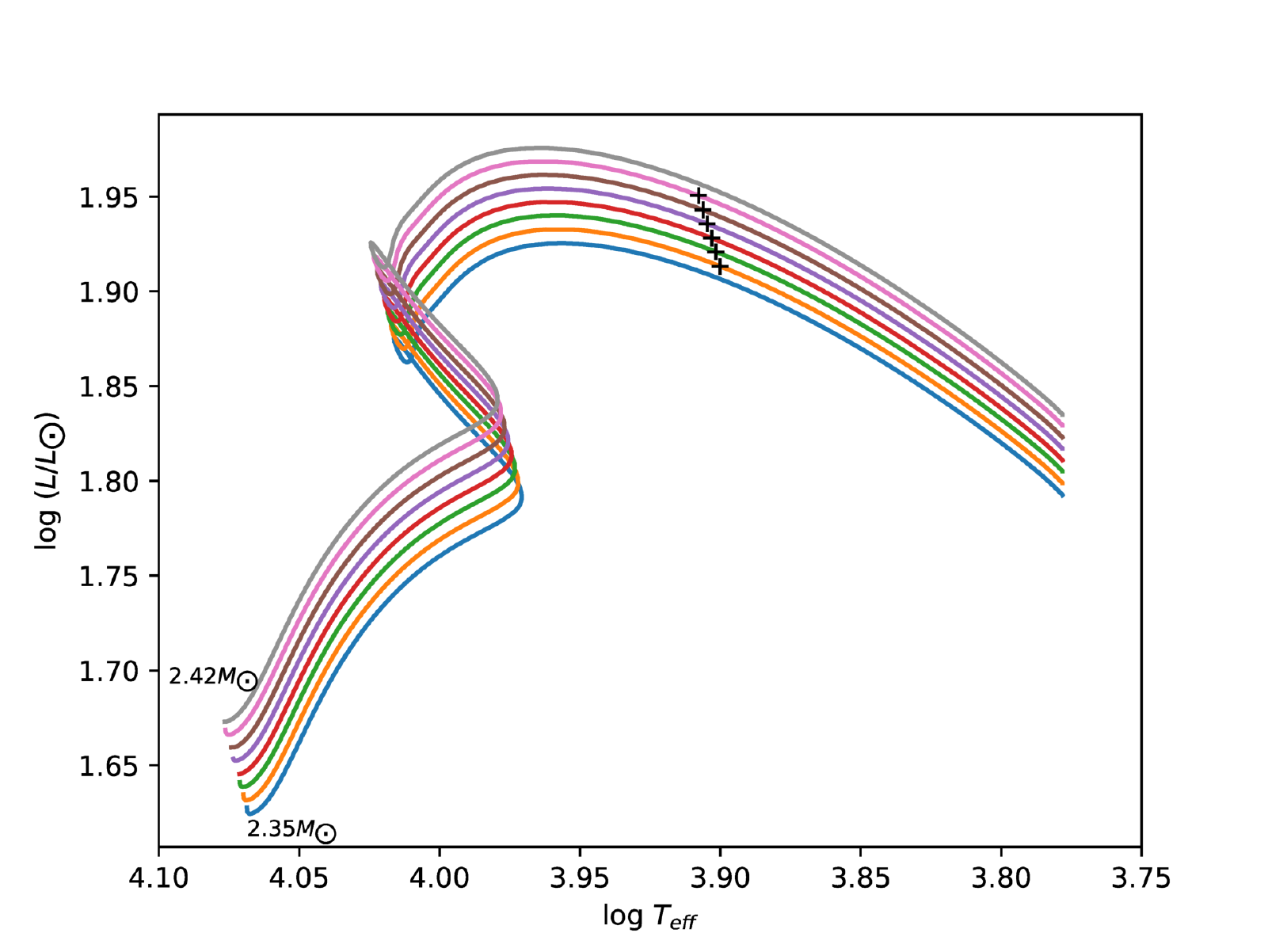}
\caption{The evolutionary tracks from the main sequence to the end of the post-main sequence in the mass range of 2.35-2.42 $M_{\odot}$. The black crosses mark the models for which the calculated $f_0$, $f_1$ and $(1/P_{0})(dP_{0}/dt)$ fit simultaneously the observation-determined  $f_{0}$, $f_{1}$ and $(1/P_{0})(dP_{0}/dt)$ within $3\sigma$.}
\label{fig:last_result}
\end{figure*}

\setcounter{table}{10}
\begin{table}
\caption{Physical Parameters of VX\ Hya Derived from the Fitted Models.}
\label{tab:best_par}
\centering
\begin{tabular}{cc}
\hline
\hline
$M$ $(M_{\odot})$      & $2.385\pm0.025$ \\
Age ($10^{8}\ \mathrm{years}$)     & $4.43\pm0.13$ \\
$T_\mathrm{eff}$ (K)& $8015\pm72$ \\
$\log(L/L_{\odot})$            & $1.93\pm0.02$ \\
$\log g$ & $3.453\pm0.001$ \\
\hline
\end{tabular}
\end{table}

\section{DISCUSSION}
\begin{figure*}
\centering
\includegraphics[width=0.7\textwidth]{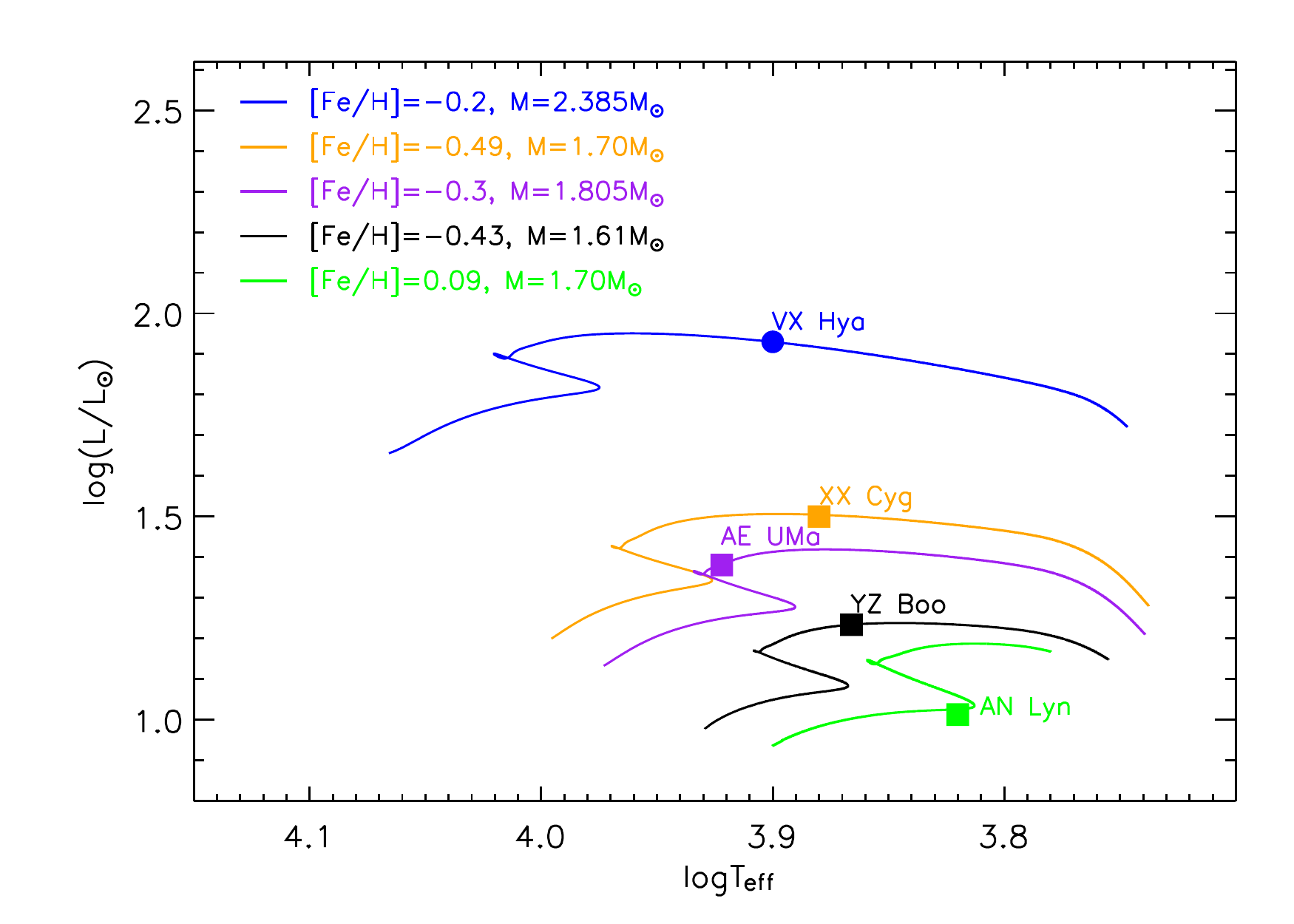}
\caption{Evolutionary tracks for the best-fit models of five HADS stars. See Table~\ref{parameters_five_stars} for details.}
\label{hr2}
\end{figure*}

\begin{table*}
\caption{Observation-determined Fundamental Frequencies, Period Change Rates and Physical Parameters from the Best-fit Models of Five HADS Stars. }
\label{parameters_five_stars}
\centering
\begin{tabular}{cccccccc}
\hline
\hline
star & $f_{0}$ ($\rm{c\ days^{-1}}$) & $(1/P_{0})(dP_{0}/dt)$ $(\mathrm{yr^{-1}})$ & $M/M_{\odot}$ & Age ($10^{9}\ \rm{years}$)& [Fe/H] &References  \\
\hline
VX\ Hya& 4.4763 & $1.81(\pm0.09)\times10^{-7}$ & 2.385 & 0.43 & -0.2  & this work \\
XX\ Cyg& 7.4148 & $1.19(\pm0.13)\times10^{-8}$& 1.70 & 0.9   & -0.49 & \citet{Yang2012} \\
YZ\ Boo& 9.6069 & $6.7(\pm0.9)\times10^{-9}$ & 1.61  & 1.44  & -0.43 & \citet{Yang2018} \\
AN\ Lyn& 10.1721& -                          & 1.70  & 1.33  & 0.09  & \citet{Li2018} \\
AE\ UMa& 11.6256& $5.4(\pm1.9)\times10^{-9}$ & 1.805 & 1.055 & -0.3  & \citet{Niu2017} \\
\hline
\end{tabular}
\end{table*}

In Figure \ref{hr} (a) and (b), one can see that large numbers of models on the tracks have the observed values of $f_{0}$ and $f_{1}$ within uncertainties, respectively. However, as shown in Figure \ref{hr} (c), only a few models satisfy the constraints from both $f_{0}$ and $f_{1}$. On the other hand, the fitting results show that the period change of VX\ Hya can be successfully ascribed to the evolutionary effect as shown in Figure \ref{fig:last_result}. In addition, our results also indicate that VX\ Hya as a HADS is a normal star evolving in the post-main-sequence stage.

As one may note, five HADS stars have been studied with the stellar masses and evolutionary stages determined by asteroseismology in our series publications. According to \citet{Yang2012}, \citet{Niu2017}, \citet{Li2018}, \citet{Yang2018} and this work, we plot the positions of the five stars on the H$-$R diagram in Figure~\ref{hr2}. The observation-determined fundamental frequencies, the period change rates, and the physical parameters from the best-fit models are listed in Table~\ref{parameters_five_stars}. As one can see, VX\ Hya has the largest mass and evolves further than the other four stars. One can use the basic pulsation relation $P\sqrt{\bar{\rho}/\rho_{\sun}}=Q$ to roughly compare the mean density $\bar{\rho}$ of the stars, where $P$ is the period of pulsation and $Q$ the pulsation constant. Compared with the other stars listed in Table~\ref{parameters_five_stars}, VX Hya has the longest $P_0$ which leads to the smallest $\bar{\rho}$, indicating that VX\ Hya evolves into the latest stage among the five HADS.

From Figure~\ref{hr2} and Table~\ref{parameters_five_stars}, one can note that the four HADS stars that locate at the post-main-sequence stages with the hydrogen-burning shell show a tendency that the star with lower fundamental frequency shows higher period change rate and evolves into later stage. Figure~5 of \citet{Breger1998} shows the theoretically calculated periods of the radial fundamental modes and their changes during late main-sequence and post-main-sequence evolution of the 1.8$M_\odot$ model, which is consistent with the tendency derived from the observations of these four stars. Based on our test, the tendency is insensitive to the mass and the metallicity of a $\delta$ Scuti star. Hence, one can roughly deduce the evolutionary stage of a $\delta$ Scuti star by its observation-determined frequency of the radial fundamental mode and period change rate.

\section{CONCLUSIONS}

By analyzing the photometric data gathered during 15 nights in 2015 of VX\ Hya, we have detected 25 frequencies that include the two independent frequencies $f_{0} = 4.4763\ \mathrm{c\ days^{-1}}$ and $f_{1} = 5.7897 \ \mathrm{c\ days^{-1}}$ and 23 harmonics and linear combinations. Based on the $P_{1}/P_{0}$ ratio, $f_{0}$ and $f_{1}$ were found to be the fundamental and the first overtone radial pulsation modes, respectively. From the results of the high-resolution spectroscopic observations, we conclude that VX\ Hya is a slow rotator with $v\sin{i} = 6.6\pm0.2\ \rm{km\ s^{-1}}$, and derive the metallicity $\rm{[Fe/H]}=-0.2\pm0.1$.

The Fourier-phase diagram method was performed to estimate the period change rates of VX\ Hya. Analysis of the collected time-series photometric data spanning over 60 years leads to determination of the period change rate of the fundamental mode as $(1/P_{0})(dP_{0}/dt) = (1.81 \pm 0.09)\times 10^{-7}\ \mathrm{yr^{-1}}$ while the constraint on the first overtone mode is weak.

 The stellar evolutionary models were constructed with the initial masses between 1.80\ $M_{\odot}$ and 2.80\ $M_{\odot}$, and $Z$ of 0.010. With the constraints from the observed values of $f_{0}$, $f_{1}$, and $(1/P_{0})(dP_{0}/dt)$ are within $3\sigma$ deviations, the stellar parameters of VX\ Hya can be determined (more details are found in Table \ref{tab:best_par}). Consequently, VX\ Hya is a HADS star lying after the second turn-off of the evolutionary track leaving the main sequence with a helium core and a hydrogen-burning shell.

 Moreover, we would like to point out that (i) for VX\ Hya, the period change can be successfully interpreted by the evolutionary effect; (ii) for VX\ Hya, the frequencies of the fundamental and the first-overtone modes could be used to determine the stellar parameters efficiently, which provided a successful example of asteroseismology on such kind of stars; (iii) the results provide a direct support to the general consensus that HADS are probably normal stars evolving in either the main-sequence or the post-main-sequence stages; (iv) the comparison of five HADS stars indicates that the frequency of the radial fundamental mode and its period change rate are sensitive to the evolutionary stages of the HADS stars.

\section*{Acknowledgments}

J.N.F. acknowledges the support from the National Natural Science Foundation of China (NSFC) through the grant 11673003 and the National Basic Research Program of China (973 Program 2014CB845700). L.F.M. and R.M. acknowledge the financial support from the DGAPA, Universidad Nacional Aut\'onoma de M\'exico (UNAM), under grant PAPIIT IN 100918. J.S. acknowledges the support from the China Postdoctoral Science Foundation (grant No. 2015M570960) and the foundation of Key Laboratory for the Structure and Evolution of Celestial Objects, Chinese Academy of Sciences (grant No. OP201406). J.S.N. acknowledges the support from the Projects 11475238 and 11647601 supported by National Science Foundation of China, and by Key Research Program of Frontier Sciences, Chinese Academy of Sciences. Special thanks are given to the technical staff and night assistants at the Yunnan Astronomy Observatory, Sierra San Pedro M\'artir observatory and the Apache Point Observatory for facilitating and helping making the observations. We acknowledge with thanks the variable star observations from the AAVSO International Database contributed by observers worldwide and used in this research. We thank Michel Bonnardeau who as an observer of AAVSO, provided us with the observations from 2011 to 2014.

\software{IRAF \citep{Tody1986, Tody1993}, PARSEC \citep{Bressan2012}, MAFAGS-OS \citet{Grupp2009}, Period04 \citep{Lenz2005}, MESA \citep{Paxton2011, Paxton2013, Paxton2015}, ADIPLS \citep{Christensen-Dalsgaard2008}}

\appendix

\center{
Detailed Information about Atomic Parameters and Fitted Models.
}

\setcounter{table}{3}
\startlongtable
\begin{deluxetable}{ccccc}
\centering
\tablecaption{Atomic Parameters of the Adopted Iron Lines and Equivalent Widths (EW) for VX Hya. \label{tab:atomic}}
\tablehead{
\colhead{$\lambda$ (\AA)} & \colhead{Species} & \colhead{$\chi$ (eV)} & \colhead{log$gf$} & \colhead{EW (m\AA)}
}
\startdata
5109.652 & Fe\,\uppercase\expandafter{\romannumeral1} & 4.30 &  -0.620 &   38.548 \\
5121.639 & Fe\,\uppercase\expandafter{\romannumeral1} & 4.28 &  -0.690 &   43.214 \\
5141.739 & Fe\,\uppercase\expandafter{\romannumeral1} & 2.42 &  -2.044 &   36.158 \\
5195.472 & Fe\,\uppercase\expandafter{\romannumeral1} & 4.22 &  -0.106 &   68.988 \\
5198.717 & Fe\,\uppercase\expandafter{\romannumeral1} & 2.22 &  -2.155 &   54.076 \\
5217.389 & Fe\,\uppercase\expandafter{\romannumeral1} & 3.21 &  -1.060 &   66.382 \\
5228.376 & Fe\,\uppercase\expandafter{\romannumeral1} & 4.22 &  -1.030 &   26.003 \\
5242.497 & Fe\,\uppercase\expandafter{\romannumeral1} & 3.63 &  -0.827 &   56.451 \\
5250.646 & Fe\,\uppercase\expandafter{\romannumeral1} & 2.20 &  -1.981 &   56.352 \\
5263.306 & Fe\,\uppercase\expandafter{\romannumeral1} & 3.27 &  -0.899 &   71.002 \\
5445.042 & Fe\,\uppercase\expandafter{\romannumeral1} & 4.39 &   0.040 &   87.637 \\
5576.096 & Fe\,\uppercase\expandafter{\romannumeral1} & 3.43 &  -0.870 &   74.376 \\
5624.542 & Fe\,\uppercase\expandafter{\romannumeral1} & 3.42 &  -0.755 &   89.210 \\
6024.058 & Fe\,\uppercase\expandafter{\romannumeral1} & 4.55 &   0.050 &   73.064 \\
6065.492 & Fe\,\uppercase\expandafter{\romannumeral1} & 2.61 &  -1.460 &   82.689 \\
6136.615 & Fe\,\uppercase\expandafter{\romannumeral1} & 2.45 &  -1.400 &   98.870 \\
6252.555 & Fe\,\uppercase\expandafter{\romannumeral1} & 2.40 &  -1.607 &   77.593 \\
6393.612 & Fe\,\uppercase\expandafter{\romannumeral1} & 2.43 &  -1.430 &   79.191 \\
6411.649 & Fe\,\uppercase\expandafter{\romannumeral1} & 3.65 &  -0.600 &   80.196 \\
6677.987 & Fe\,\uppercase\expandafter{\romannumeral1} & 2.69 &  -1.318 &   87.169 \\
5983.690 & Fe\,\uppercase\expandafter{\romannumeral1} & 4.55 &  -0.538 &   26.675 \\
6003.002 & Fe\,\uppercase\expandafter{\romannumeral1} & 3.88 &  -0.990 &   37.281 \\
6056.010 & Fe\,\uppercase\expandafter{\romannumeral1} & 4.73 &  -0.360 &   39.274 \\
6078.490 & Fe\,\uppercase\expandafter{\romannumeral1} & 4.79 &  -0.171 &   35.711 \\
6191.571 & Fe\,\uppercase\expandafter{\romannumeral1} & 2.43 &  -1.417 &   99.623 \\
6246.327 & Fe\,\uppercase\expandafter{\romannumeral1} & 3.60 &  -0.773 &   64.747 \\
6301.508 & Fe\,\uppercase\expandafter{\romannumeral1} & 3.65 &  -0.718 &   71.379 \\
6302.499 & Fe\,\uppercase\expandafter{\romannumeral1} & 3.69 &  -0.973 &   55.055 \\
6430.856 & Fe\,\uppercase\expandafter{\romannumeral1} & 2.18 &  -1.966 &   70.720 \\
5132.669 & Fe\,\uppercase\expandafter{\romannumeral2} & 2.81 &  -4.100 &   35.766 \\
5264.808 & Fe\,\uppercase\expandafter{\romannumeral2} & 3.23 &  -3.050 &   90.487 \\
5425.257 & Fe\,\uppercase\expandafter{\romannumeral2} & 3.20 &  -3.280 &   70.668 \\
5991.380 & Fe\,\uppercase\expandafter{\romannumeral2} & 3.15 &  -3.600 &   56.818 \\
6084.111 & Fe\,\uppercase\expandafter{\romannumeral2} & 3.20 &  -3.831 &   35.442 \\
6416.919 & Fe\,\uppercase\expandafter{\romannumeral2} & 3.89 &  -2.977 &   66.339 \\
6432.683 & Fe\,\uppercase\expandafter{\romannumeral2} & 2.89 &  -3.610 &   71.063 \\
6516.080 & Fe\,\uppercase\expandafter{\romannumeral2} & 2.89 &  -3.272 &  106.256 \\
\enddata
\tablecomments{$\chi$ is the excitation energy. log$gf$ is the oscillator strengths.}
\end{deluxetable}
\clearpage

\setcounter{table}{9}
\begin{table}\footnotesize
\caption{Models within 3$\sigma$ Deviations of $f_{0}$, $f_{1}$ and $(1/P_{0})(dP_{0}/dt)$.}
\label{tab:fitting_results}
\begin{center}
\begin{tabular}{ccccccccc}
\hline
\hline
$M$ $(M_{\odot})$ &Age ($10^{8}\ \mathrm{years}$) &$\log T_\mathrm{eff}$ &$\log(L/L_{\odot})$   &$\log g$ &$f_{0}$ ($\rm{c\ days^{-1}}$) &$f_{1}$ ($\rm{c\ days^{-1}}$) &$\frac{1}{P_{0}}\frac{\rm{d}P_{0}}{\rm{d}t}$ ($\times 10 ^{-7} \text{yr}^{-1}$) &$\chi^{2}$ \\
\hline
2.36 &4.55 &3.9001 &1.9132 &3.4513 &4.4754 &5.7910 &1.54 &21.11  \\
\hline
2.37 &4.50 &3.9017 &1.9209 &3.4519 &4.4761 &5.7909 &1.56 &13.69  \\
2.37 &4.50 &3.9016 &1.9209 &3.4518 &4.4757 &5.7904 &1.56 &12.06  \\
2.37 &4.50 &3.9016 &1.9209 &3.4518 &4.4753 &5.7899 &1.56 &14.36  \\
\hline
2.38 &4.45 &3.9032 &1.9281 &3.4525 &4.4769 &5.7909 &1.58 &15.03 \\
2.38 &4.45 &3.9032 &1.9281 &3.4524 &4.4765 &5.7904 &1.58 &8.54  \\
2.38 &4.45 &3.9031 &1.9281 &3.4524 &4.4760 &5.7898 &1.58 &7.38  \\
2.38 &4.45 &3.9031 &1.9281 &3.4523 &4.4756 &5.7893 &1.58 &10.01 \\
2.38 &4.45 &3.9031 &1.9281 &3.4523 &4.4753 &5.7889 &1.58 &16.43 \\
\hline
2.39 &4.40 &3.9047 &1.9356 &3.4531 &4.4771 &5.7903 &1.60 &11.21  \\
2.39 &4.40 &3.9047 &1.9356 &3.4530 &4.4768 &5.7898 &1.59 &7.21  \\
2.39 &4.40 &3.9047 &1.9356 &3.4530 &4.4764 &5.7893 &1.59 &6.41  \\
2.39 &4.40 &3.9047 &1.9356 &3.4529 &4.4761 &5.7889 &1.59 &8.62  \\
2.39 &4.40 &3.9047 &1.9356 &3.4529 &4.4757 &5.7885 &1.60 &13.99  \\
\hline
2.40 &4.35 &3.9062 &1.9430 &3.4536 &4.4772 &5.7894 &1.61 &10.47  \\
2.40 &4.35 &3.9062 &1.9429 &3.4535 &4.4768 &5.7888 &1.61 &9.28  \\
2.40 &4.35 &3.9062 &1.9429 &3.4535 &4.4764 &5.7883 &1.62 &12.11 \\
\hline
2.41 &4.31 &3.9078 &1.9506 &3.4541 &4.4773 &5.7885 &1.63 &16.00  \\
\hline
\end{tabular}
\end{center}
\end{table}


\end{document}